\documentclass[aps,prb,twocolumn,superscriptaddress,footinbib,notitlepage,floatfix]{revtex4-2}
\usepackage{amssymb}
\usepackage{amsmath}
\usepackage{mathtools}
\usepackage{mathdots}
\usepackage[normalem]{ulem}
\usepackage{txfonts}
\usepackage{array}
\usepackage{amssymb}
\usepackage{amsfonts}
\usepackage{amsmath}
\usepackage{mathrsfs}
\usepackage{booktabs}
\usepackage{threeparttable}
\usepackage{multirow}
\usepackage[dvips]{graphicx}
\usepackage{epsfig}
\usepackage{graphicx}
\usepackage{subfigure}
\usepackage{threeparttable}
\usepackage{chngpage}
\usepackage{float}
\usepackage[dvipsnames,svgnames,x11names,hyperref]{xcolor}
\usepackage{bm}
\usepackage[toc]{appendix}
\usepackage{tabularx}
\usepackage{adjustbox}
\usepackage{comment}
\usepackage{hyperref}
\definecolor{lightpurple}{RGB}{200, 160, 220} 
\usepackage{standalone}
\usepackage{bbding}
\hypersetup{
	colorlinks=true,
	linkcolor=NavyBlue,
	citecolor=blue, 
	filecolor=magenta,
	urlcolor=blue
}

\newcommand{\bs}{{\boldsymbol{s}}}

\usepackage{upgreek}

\begin{document}
	
	\title{Analytical quantification of strongly disordered discrete time crystals}
	\author{Yang-Ren Liu}
	\affiliation{Kavli Institute for Theoretical Sciences, University of Chinese Academy of Sciences, Beijing 100190, China}
	\author{Biao Huang}
	\email[Corresponding author: ]{phys.huang.biao@gmail.com}
	\affiliation{Kavli Institute for Theoretical Sciences, University of Chinese Academy of Sciences, Beijing 100190, China}
	
	\date{\today}

	\begin{abstract}
		Discrete time crystals (DTCs) have emerged as a novel non-equilibrium phenomenon, exhibiting spontaneous symmetry breaking of discrete time translation symmetry, and holding promise for applications ranging from enhancing quantum sensing to protecting quantum entanglement. 
		Nevertheless, analytical quantification for disordered DTCs remains challenging, leaving uncertainties about their behaviors in large scale systems amid current debates on the stability of (Floquet-)many-body localization. In this work, we introduce an analytical framework to calculate the values of key observables in a strongly disordered DTC without fitting parameter. The perturbatively obtained closed-form formulae show quantitative agreement with numerical simulations for inverse participation ratios for eigenstate localization in Fock space, Edwards-Anderson parameters for spin-glass orders, mutual information for long-range entanglement, and the steady-state amplitudes of autocorrelators for period-doubled oscillations. Meanwhile, we demonstrate that eigenstate resonances render the scaling for the deviation of physical observables from their unperturbed values as $O(\lambda)$, in contrast to non-resonant situations with suppressed deviation $O(\lambda^2)$. Our newly developed scheme adopts resolvent perturbation formalism, which can directly prescribe arbitrarily higher-order corrections without iterations. With such advantages, we analytically prove that quasienergy corrections for pairwise cat eigenstates are identical up to order $O(\lambda^{(L/n_{\text{op}})-1})$, where perturbations of strength $\lambda$ involve at most $n_{\text{op}}$-spin terms. Such spectral pairing deviations quantify the DTC lifetime as $\tau_* \sim (1/\lambda)^{L/n_{\text{op}}}$. We further show that the analytical framework applies to generic DTC models with dominant Ising interaction and a given number of qubits, offering a unified scheme to quantify physical observables for systems beyond numerically accessible sizes, with or without disorders.

	\end{abstract}
	\maketitle 
	
	\section{Introduction}\label{sec1}

	Discrete time crystals (DTCs)~\cite{sacha2017RPPa,khemani2019a,else2020ARCMPa,zaletel2023RMPa} have emerged as an intriguing state of matter in far-from-equilibrium systems. They exhibit rigid subharmonic responses for physical observables compared with the underlying driving frequency, which is robust against arbitrary perturbation. One major, and prototypical~\cite{khemani2016PRLa,else2016PRLb,yao2017PRLa,vonkeyserlingk2016PRBa,ippoliti2021PQ}, paradigm to stabilize such a system exploited strong quenched disorders to contain runaway heating~\cite{dalessio2014PRX,lazarides2014PRE}, a phenomenon known as many-body localization (MBL)~\cite{sierant2025RPP}. Researches over the first decade, initiated with the analytical~\cite{Basko2006} and numerical~\cite{oganesyan2007PRBa} works, proposed the scheme~\cite{nandkishore2015ARCMPa,alet2018CRPa,abanin2019RMPb} featuring a sharp localization phase transition at a certain critical disorder strength. While the analytical frameworks for low-temperature diagrammatic perturbation theory~\cite{Basko2006} and full diagonalization by Jacobi decimation methods~\cite{imbrie2016JSP,imbrie2016PRL} are designed only for static systems~\cite{zaletel2023RMPa}, numerics suggests a possible localization transition also in periodically driven Floquet systems~\cite{abanin2016AoPa, ponte2015AoP,ponte2015PRLa,lazarides2015PRLa,bordia2017NP}. Inspired by such successes, numerous mechanisms other than MBL have also been proposed to stabilize DTC responses for selected initial states, including Floquet prethermalization \cite{else2017PRXb,machado2020PRXa,ho2023AP,luitz2020PRX,stasiuk2023PRX,abanin2017PRB,peng2021NP,yokota2025a}, quantum many-body scars \cite{Maskara2021PRL,huang2022PRLa,huang2023PRBa,bull2022PRL,deng2023PRB}, and others \cite{russomanno2017PRB,chinzei2020PRL,gong2018PRL,pizzi2021NC,zhu2019NJP,liu2023PRL,kumar2025PRB,collura2022PRX}. Experimental realizations of DTCs have appeared in wide ranges of platforms~\cite{choi2017N,zhang2017N,Rovny2018,frey2022SA,randall2021S,mi2022N,Kyprianidis2021,Beatrez2023,Stasiuk2023,bao2024NCa,Kongkhambut2022,Wu2024}.

	In the past decade, however, studies have pointed out a mechanism devastating for MBL itself, dubbed quantum avalanche effects. Initial works~\cite{deroeck2017PRBa,luitz2017PRL,thiery2018PRLa} showed the instability of higher dimensional MBL~\cite{Li2025} to rare (Griffiths) regions with weak disorders, which serve as thermal bubbles to trigger avalanching delocalization. After 2019, it was further discovered, both numerically~\cite{sels2021PREa,sels2022PRBa,suntajs2020PRBa,suntajs2020PREa,morningstar2022PRBa,ha2023PRLa,Long2023} and experimentally~\cite{Leonard2023}, that even one-dimensional disordered systems with short-range interaction --- a scenario previously thought to be most stable for MBL --- fail to maintain localization when one takes the large system size and long-time limit. The evidence for such an instability comes from several aspects. For instance, the critical disorder strength for MBL phase transition extracted numerically keeps drifting towards the strong disorder limit with increasing system sizes, a finite-size effect indicating that at thermodynamic limits, MBL may cease to exist at any finite disorder strength. Also, simulations of MBL systems coupled to thermal bubbles indicate that one needs a disorder strength about one order of magnitude stronger than what was found previously (i.e., by level spacing statistics) to avoid thermalization by quantum avalanche.
	On the other hand, available analytical theory does not appear suitable to calculate the specific values of physical observables in MBL systems, as the predicted quantities typically involve fitting parameters to be determined numerically. That brings about uncertainty depending on different numerical methods and interpretations. For instance, assuming an MBL phase transition, exact diagonalization gives a critical exponent $\nu\sim 1$ ~\cite{Khemani2017,Luitz2015,Kjaell2014} violating Harris bound $\nu\geqslant 2$~\cite{Harris1974} in one-dimension, renormalization group methods give instead $\nu\sim 3$~\cite{Vosk2015,Zhang2016,Potter2015,Dumitrescu2017}, and more recently, exact diagonalization from the Fock space perspective predicts a Berezinskii-Kosterlitz-Thouless type of non-polynomial scaling~\cite{Laflorencie2020,Dumitrescu2019,Goremykina2019,suntajs2020PRBa}.
	Currently, debates continue as for whether a well-defined MBL phase, in the sense of having a finite disorder strength for localization-thermal phase transition, can exist deep inside the strongly disordered regime~\cite{Abanin2021,kiefer-emmanouilidis2020PRLa,kiefer-emmanouilidis2021PRBa,Sels2023,Colbois2024,crowley2022SPa}. Meanwhile, it is suggested that a ``prethermal MBL"~\cite{Long2023,morningstar2022PRBa} crossover occurs in the regime with relatively weak disorder strength, replacing the role of MBL phase transition in earlier literature.

	The uncertainty in underlying MBL mechanism presents a nontrivial obstacle to researches on MBL-based DTCs. Indeed, the preexisting MBL-DTC theories~\cite{khemani2016PRLa,else2016PRLb,yao2017PRLa} all {\em assume} robust Floquet-MBL and its resulting structures, such as a complete set of local integrals of motions (LIOM), based on which the DTC theory is further constructed. Such constructions are pending full justification as a mathematical proof for the prerequisite Floquet-MBL itself is still absent~\cite{zaletel2023RMPa}. Nevertheless, the latest development in quantum computing devices has suggested an unconventional but useful scheme shift: rather than asking about whether Floquet-MBL phase transitions exist in the thermodynamic limit (i.e. at system size $L\rightarrow\infty$), which was the focus of previous theories, it is practically more relevant to quantify physical observables in a finite-size system with a large number of qubits $\infty > L\gg1$.     
	Specifically, current quantum devices, such as the superconducting qubits~\cite{Kim2023,Acharya2024,Jin2025,Gao2025} and Rydberg atoms~\cite{Bluvstein2023,Evered2025}, have realized digital circuits containing $L>100$ qubits, which is far from the thermodynamic limit but also well beyond the size that numerics can fully diagnose (i.e., $L\lesssim 18$). It raises questions of how to faithfully relate numerical findings to experimental observations on quantum devices with one order of magnitude more qubits, given the notable drift of critical disorder strength for MBL transition or crossover with system sizes. Aside from scientific curiosity, there are also practical demands to quantify physical observables on large-scale DTCs targeting technological applications, such as protecting Greenberger-Horne-Zeilinger (GHZ) entanglement~\cite{bao2024NCa} and enhancing quantum sensing~\cite{Moon2024,Yousefjani2025,Iemini2024}.

	To explore the possible solutions, in this work, we construct an analytical theory to calculate the values of physical observables for DTCs in the strongly disordered regime. We take a generalized Fock space perspective, where each idealized GHZ state is regarded as a lattice site, while perturbations to the Floquet unitary generate connections among different sites. Based on this, we generalize the resolvent formalism~\cite{kato1949PoTP,messiah1999quantum}, originally designed for static quantum mechanical problems, to a strongly interacting Floquet many-body system. Compared with usual perturbation theory that requires an order-by-order iteration, the resolvent formalism can directly give arbitrary higher-order corrections. This offers tremendous simplification to prove the robustness of spectral-pairing essential to the stability of DTC oscillation frequency, which requires analysis of up to $L$-th order terms in the perturbations series.
	
	Our results include two major parts. First, in the strongly disordered regime, the perturbation theory gives the explicit form of the first-order corrected eigenstates, which then further prescribes the values of key physical observables characterizing a long-range-entangled DTC. They include the eigenstate inverse participation ratio (IPR), the Edwards-Anderson parameter, mutual information, and the amplitude of autocorrelators in steady states. Second, our framework gives formal expressions of higher-order quasienergy corrections, which not only qualitatively proves the spectral-pairing rigidity but also quantitatively predicts the values of scaling exponents. Specifically, an ideal (unperturbed) DTC would host pairwise cat eigenstates separated by quasienergy gap $\pi$ for the period-doubled oscillation with infinite lifetime. Such a gap receives an exponentially small correction $\pi + O(e^{-\kappa L})$ in a finite system of $L$ spins, thereby giving an exponentially long lifetime $\sim e^{\kappa L}$ for DTC oscillations. Our theory predicts that the exponent $\kappa$ can be simply determined by checking the form of perturbation, i.e., $\kappa=1$ for single spin perturbations $\sim \sigma^x_j$, $\kappa=1/2$ for two-spin perturbation $\sim \sigma^x_j + \sigma^x_j \sigma^x_{j+1}$, and etc. 
	We benchmark aforementioned analytical predictions against numerical results, and find a quantitative agreement in the strongly disordered regime without any fitting parameters. The framework is applicable to generic DTC models with dominant Ising interaction, with or without disorders, as we demonstrate with examples. Such a framework supplies generic tools for future researches to predict new phenomena and quantify possible technological applications, and offers independent benchmark of experimental results for large-scale quantum devices.
	
	It is worth clarifying the relation of our analytical framework with previous works on interacting localization problems viewed from Fock space perspectives. The parameter regime we consider resides deeply in the strongly disordered side, such that each Floquet eigenstate is dominated by an $O(1)$ number of Fock bases. Such a concept of Fock space localization originates from the work by Altshuler~\cite{Altshuler1997} in the studies of quantum dots, with the difference that we consider the correlated Fock space network corresponding to spin chains, instead of an uncorrelated Cayley tree network in Ref.~\cite{Altshuler1997}. 
	Also, Fock space localization here should be distinguished from the relatively weakly disordered case with multifractal dimensions~\cite{mace2019PRL}, where eigenstates are delocalized in Fock space, occupying exponentially many Fock bases with distributions different from a thermal one. Technically, the Fock space localized regime with IPR$\rightarrow 1$ would exhibit quantum dimension $-\ln(\text{IPR})/\ln(2^L) \rightarrow 0$, which can be thought of as an Anderson localization in Fock space network. This is in contrast to the thermal regime with dimension $1$ and the multifractal regime sandwiched in between. 
	
	Aside from providing practical tools in the study of large-scale quantum devices, we are motivated to focus on such an unusual regime with strong disorders also to pave the way for analytical understanding of avalanche effects, which are triggered exactly in such a regime~\cite{sels2022PRBa,morningstar2022PRBa}. Indeed, available analytical theories~\cite{Basko2006,imbrie2016JSP,imbrie2016PRL} focus chiefly on the weakly disordered regime, i.e., with multifractal dimensions. More importantly, as emphasized in Ref.~\cite{morningstar2022PRBa}, the rarely explored regime deep inside the localized side is far from being trivial, but exhibits rich landscape of resonances that deserves further investigations. This is verified by our theory quantifying that resonances among pairwise cat eigenstates proliferate in such a strongly disordered regime and change the scaling behavior of physical observables. For instance, IPRs of majority eigenstates in disordered DTCs deviate from the idealized value as a function of $O(\lambda L)$ as a result of pairwise resonances, where $\lambda\ll1$ is the perturbation strength. This is to be contrasted against the non-resonant cat scars~\cite{huang2023PRBa,bao2024NCa} in translation-invariant limit, where the IPR is closer to the idealized value with a suppressed deviation that scales as $O(\lambda^2L)$. 
	
	The rest of the paper is organized as follows. For intuitiveness, in Sec.~\ref{sec2}, we illustrate the general framework and its applications to a concrete example, where we summarize the explicit analytical form of physical observables and compare them with numerical data. To assist future works, we exemplify further applications of our analytical theory to additional models in Sec.~\ref{sec4}, including more generic disordered cases with one-spin and two-spin perturbations together with a contrast against translation-invariant cases with cat scars. The details and general framework of analytical theory are furnished in Appendix~\ref{sec3}.  We finally summarize in Sec.~\ref{sec5}.

	\section{Illustration of the results}\label{sec2}
	
	In this section, we illustrate the analytical framework and its application to a specific model, showing the agreement between analytical predictions with numerical results. The full derivation, which appears a bit lengthy, is left to Appendix~\ref{sec3}. These discussions offer a roadmap to more generic applications as shown in later sections.
	
	\subsection{General description of the analytical framework}
	
	We study a class of strongly disordered models with the Floquet operator for evolution over a period of the form $U_F = U_0e^{i\lambda V}$. Here, the unperturbed part $U_0=\sum_\alpha e^{i E_\alpha}|E_\alpha\rangle\langle E_\alpha|$ is diagonal in Fock bases (up to a perfect global spin flip), yielding exactly solvable eigenstates and eigenvalues as $|E_\alpha\rangle$ and $E_\alpha$, respectively. $U_0$ involves terms like the random Ising interaction and longitudinal fields. These disordered diagonal terms enforce the Fock space localization. Then, $V$ represents a generic local perturbation off-diagonal in Fock bases, such as the transverse fields and transverse interactions, which tend to delocalize the system by mixing different Fock states. The perturbation strength is characterized by $\lambda$. Our goal is to perturbatively diagonalize the full Floquet operator $U_F = \sum_\alpha e^{i\omega_\alpha}|\omega_\alpha\rangle \langle \omega_\alpha| $ under weak perturbation $\lambda\ll1$, based on which we extract the values for physical observables.

	It is worth emphasizing that we do {\em not} assume the existence of MBL {\em a priori}; instead, our framework targets accounting for both the localization and DTC features simultaneously. 
	Meanwhile, the long-range entangled DTC property imposes peculiar requirements absent from static MBL systems.
	Thus, we are led to abandon previous MBL theories and construct a new framework from scratch to accommodate these demands, as specified below.

	First, the theory should be able to directly prescribe quasienergy corrections of arbitrarily higher-orders. This is rarely a requirement for any previous perturbation theories, as the ($n$-th) higher-order correction is exponentially small $\lambda^n, \lambda\ll1$, so in most scenarios only the first few orders are needed. However, the central character for DTCs is that its oscillation lifetime $\sim e^L$ extends exponentially with system size $L$, reflecting the long-range entanglement of underlying cat eigenstates. As we shall show, this arises from that quasienergy corrections are identical for pairwise cat eigenstates all the way until $\propto L$-th order corrections. Thus, we do not adopt conventional perturbation formalism in static MBL systems~\cite{Basko2006,imbrie2016JSP,imbrie2016PRL} that assesses corrections order-by-order with iterations.
	
	Instead, we construct a Floquet-resolvent formalism for such a task (see Appendix~\ref{sec4-1} for more detailed algebras). In this framework, we define two resolvents $G(z)$ and $G_0(z)$ corresponding to the perturbed $U_F$ and unperturbed $U_0$ evolution operator, respectively,
	\begin{align}\label{eq:resolvent}
		G(z)=\frac{1}{z-U_F},
		\qquad
		G_0(z) = \frac{1}{z-U_0}.
	\end{align}
	They can be related by a Floquet-Born series,
	\begin{align}\label{eq:born}
		G = G_0(U_F-U_0)G = \sum_{n=0}^\infty G_0 [(U_F-U_0)G_0]^n.
	\end{align}
	Note that $U_F-U_0 = U_0(e^{i\lambda V}-1) = O(\lambda)$, the $n$-th term is of correction order $O(\lambda^n)$.
	Meanwhile, with complex-valued $z$, we can perform a contour integral $\mathscr{C}(\alpha)$ around a certain quasienergy $e^{i\omega_\alpha}$ for the resolvent. Using residual theorem and the spectral representation $G(z)=\sum_\alpha |\omega_\alpha\rangle (z-e^{i\omega_\alpha})^{-1} \langle\omega_\alpha|$, such a contour integral produces the projection operator onto the eigenstate $|\omega_\alpha\rangle$,
	\begin{align}
		P(\alpha) = \oint_{\mathscr{C}(\alpha)} \frac{dz}{2\pi i}\, G(z) = |\omega_\alpha\rangle \langle\omega_\alpha|.
	\end{align}
	Finally, one can consider the trace of the operator $(U_F - e^{iE_\alpha})P(\alpha)$, which gives the explicit form showing the quasienergy correction in $\omega_\alpha$ with respect to $E_\alpha$ of arbitrary orders,
	\begin{align}\nonumber
		&
		\quad e^{i\omega_\alpha} - e^{iE_\alpha} 
		\\
		\label{eq:qe_correction}
		&= 
		\sum_{n=1}^\infty \frac{1}{2\pi i}
		\oint_{{\mathscr C}(\alpha)}
		\left(z-e^{iE_\alpha}\right) G_0(z) \left[U_0 \left(e^{i\lambda V}-1 \right) G_0(z)\right]^n 
		\mathrm{d}z.
	\end{align}
	Here, we have used the expansion in Eq.~\eqref{eq:born}, where the index $n$ corresponds to corrections of order $O(\lambda^n)$.

	Second, our theory should also independently quantify the  stability of long-range entangled cat eigenstates under perturbations. One may write the perturbed eigenstate $|\omega_\alpha\rangle$ in terms of the unperturbed ones $|E_\alpha\rangle$ by using the projection operator,
	\begin{align}\nonumber
		|\omega_\alpha\rangle 
		& =
		\frac{1}{\sqrt{N}}P(\alpha) |E_\alpha\rangle
		\\ \label{eq:non-deg_eig}
		&=
		\frac{1}{\sqrt{N}} \left(|E_\alpha\rangle + i\lambda \sum_{\alpha'} \frac{A_{\alpha,\alpha'}}{e^{iE_\alpha} - e^{iE_{\alpha'}}} |E_{\alpha'}\rangle \right) + O(\lambda^2).
	\end{align}
	Here $N$ is a normalization constant, $A_{\alpha,\alpha'}$ involves the matrix element $\langle E_{\alpha'}|V| E_\alpha\rangle$. However, similar to any perturbation theory, the denominator of Eq.~\eqref{eq:non-deg_eig} tends to diverge if two quasienergies $E_\alpha, E_{\alpha'}$ approach degeneracy. While in a strongly disordered system, one never encounters rigorous degeneracy, we find that for arbitrarily weak perturbation strength $\lambda$, there is always a notable chance for $|A_{\alpha,\alpha'}/(e^{iE_\alpha} - e^{iE_{\alpha'}})|\gg 1/\lambda$ in Eq.~\eqref{eq:non-deg_eig} to occur, which invalidates the convergence of perturbation series.
	To remedy such a pathological behavior, one may tentatively introduce certain artificial cutoff and abandon those quasienergy pairs or disorder samples leading to small denominators for Eq.~\eqref{eq:non-deg_eig}. However, we find that the values of predicted physical observables would sensitively depend on such a cutoff, implying that the approximate degeneracy is an essential eigenstate property and cannot be removed artificially.
	
	Thus, a modified scheme taking into account degenerate hybridization is necessary, which is the second important ingredient for our theory to quantify localization in DTCs without using fitting parameters. 
	We postulate that, in the strongly disordered regime (with small $\lambda$), degeneracy is only relevant between pairwise eigenstates; that is, one can safely ignore the chances for three or more $E_{\alpha}$'s to be almost degenerate. This is based on the observation that probability distribution $P(\Delta)$ for the gap $\Delta\equiv E_\alpha-E_{\alpha'}$ is smooth. Then, the probability for a small denominator $|e^{iE_\alpha} - e^{iE_{\alpha'}}| = |e^{i(E_\alpha - E_\alpha')}-1| \approx |\Delta| <\lambda\ll1$, namely, $\int_{-\lambda}^\lambda P(\Delta)\, d\Delta$, is smooth and proportional to $\lambda$, so that more levels simultaneously being close to degeneracy corresponds to higher order corrections. Thus, truncated to the first order in $\lambda$, as in Eq.~\eqref{eq:non-deg_eig}, it is legitimate to consider pairwise eigenstate degeneracy alone. 
	
	The detailed treatment is presented in Appendix~\ref{sec3-3}, which we sketch below. First, we expand the perturbed Floquet operator $U_F\approx U_0(1+i\lambda V)$ within each subspace spanned by the target eigenstate $|E_\alpha\rangle$ and the relevant one $|E_{\alpha'}\rangle$ contributing to Eq.~\eqref{eq:non-deg_eig}, 
	\begin{align}\label{eq:mix_matrix}
		\begin{pmatrix}
			e^{iE_\alpha} & i\lambda e^{iE_{\alpha}} \langle  E_{\alpha}|V|E_{\alpha'}\rangle \\
			i\lambda e^{iE_{\alpha'}} \langle E_{\alpha'}|V|E_\alpha\rangle & e^{iE_{\alpha'}}
		\end{pmatrix}.
	\end{align}
	Diagonalizing it, we formally obtain the eigenstates
	\begin{align}
		|\tilde{\omega}_{\alpha,\alpha'}\rangle = u_{\alpha,\alpha'} |E_\alpha\rangle + v_{\alpha,\alpha'}|E_{\alpha'}\rangle 
		=
		u_{\alpha,\alpha'}\left(
		|E_\alpha\rangle + \frac{v_{\alpha,\alpha'}}{u_{\alpha,\alpha'}} |E_{\alpha'}\rangle
		\right).
	\end{align}
	The coefficient $v_{\alpha,\alpha'}/u_{\alpha,\alpha'}$ would reduce to the non-degenerate form $i\lambda A_{\alpha,\alpha'}/(e^{iE_\alpha} - e^{iE_{\alpha'}})$ in Eq.~\eqref{eq:non-deg_eig} in the far from degeneracy situation $|A_{\alpha,\alpha'}/(e^{iE_\alpha} - e^{iE_{\alpha'}})|\ll 1/\lambda $. Then, one can write the first-order corrected eigenstate, with or without degeneracy, in a unified form as
	\begin{align}\label{eq:deg_eigs}
		|\omega_\alpha\rangle = \frac{1}{\sqrt{N}} \left[
		|E_\alpha\rangle + \sum_{\alpha'} \frac{v_{\alpha,\alpha'}}{u_{\alpha,\alpha'}} |E_{\alpha'}\rangle
		\right] + O(\lambda^2).
	\end{align}
	This replacement is a new approximation method we develop in this work, based on the pairwise resonance postulation discussed above. 
	Namely, if two levels $(\alpha, \alpha')$ are far from resonance, their contribution to Eq.~\eqref{eq:deg_eigs} are same as that derived from non-degenerate perturbation theory in Eq.~\eqref{eq:non-deg_eig}; if resonance occurs to any two levels, Eq.~\eqref{eq:deg_eigs} reproduces the near-degenerate hybridization results. 
	Practically, Eq.~\eqref{eq:deg_eigs} enjoys the advantage that it connects the two limits with smooth functions, so one does not need to manually identify which pairs are resonant, and can efficiently compare analytical predictions with numerical or experimental results with large number of disorder realizations.
	
	The general results in Eqs.~\eqref{eq:qe_correction} and \eqref{eq:deg_eigs} characterize quasienergy and eigenstate corrections, respectively. They constitute the basis for our later discussions. 
	
	Before moving into detailed treatment of DTC physics, let us comment on the broader applicability of the new framework we introduced above. While this formalism is particularly designed to analyze DTC-related physics, it can equally be applied to static disordered systems, i.e., by replacing $U_F$ and  $U_0$ with static Hamiltonians in Eq.~\eqref{eq:resolvent} and perform similar analysis. We expect that the pairwise degeneracy, related to Eqs.~\eqref{eq:mix_matrix} -- \eqref{eq:deg_eigs}, would still play essential roles there, because this is only related to the eigenstate localization property. On the other hand, the ability to reach extremely higher-order corrections, as shown in Eq.~\eqref{eq:qe_correction}, may be an overkill for most systems irrelevant of long-range entangled DTC physics.

	\subsection{Applications to a specific model}
	Let us adopt a paradigmatic DTC model in one-dimension with short-range Ising interactions, with the Floquet operator for evolution over one period $T$ as
	\begin{align}
		\label{eq:UF_general}
		U_F&=U_0e^{i\lambda V},\quad
		U_0 =e^{-i\sum_{j=1}^L \left( J_j\sigma _{j}^{z}\sigma _{j+1}^{z}+h_j\sigma _{j}^{z} \right) }
		\prod_{j=1}^L (-i\sigma^x_j).
	\end{align}
	Here $\sigma^{x,y,z}_j$ are Pauli operators at the $j$-th site, $j=1,2,...,L$, and $\lambda$ represents imperfections of spin flips that will be stabilized by interactions. The Ising interaction $J_j$ at different sites are independent and randomly drawn from the range $[J_0-W/2,J_0+W/2]$, with $J_0=1$ throughout the paper and $W$ the disorder strength, and the longitudinal fields $h_j\in[0,2\pi]$ take maximal randomness. Periodic boundary conditions are imposed to exclude edge Majorana modes. Due to the presence of both transverse and longitudinal fields, the model in Eq.~\eqref{eq:UF_general} is generally nonintegrable and cannot be mapped to free fermion models.
	
	At $\lambda=0$, $U_0$ is exactly solvable~\cite{else2016PRLb}, with eigenstates made of pairwise cat states separated by quasienergy $\pi$. Each eigenstate is specified by the spin pattern $\boldsymbol{s}$ and the spectral-pairing quantum number $\ell$,
	\begin{align}\nonumber
		U_0|\bs,\ell\rangle&=e^{iE_{\bs,\ell}}|\bs,\ell
		\rangle,
		\quad E_{\bs,\ell}=-\sum_{j=1}^L\left( J_js_js_{j+1} \right)+\ell \pi,
		\quad
		\ell = 0,1;
		\\ \label{eq:U0_solutions_sec2}
		|\bs,\ell\rangle &=\frac{1}{\sqrt{2}}\left[ e^{-i\sum_{j=1}^L h_js_j/2}|\bs\rangle +\left( -1 \right)^\ell
		e^{i\sum_{j=1}^L h_js_j/2} \left|-\bs \right\rangle \right],
	\end{align}
	where $\bs$ denotes the Fock state
	\begin{align}
		& \left|\bs\right\rangle  = \left|s_1\right\rangle \otimes \left|s_2\right\rangle \otimes \dots \otimes \left|s_L\right\rangle,
		\qquad
		s_j = \pm1.
	\end{align}
	Our analytical framework allows for perturbatively calculating the values of physical observables in the strongly disordered regime $J_0, W\gg \lambda$ for generic perturbations $V$, which has not been achieved before.

	Let us exemplify the results with single-spin perturbations,
	\begin{align} \label{eq:UF_nop1}
		V=\sum_{j=1}^L \sigma^x_j,
	\end{align}
	while more generic cases are left to Sec.~\ref{sec4}. In the strongly disordered regime, the dominant Ising interaction ensures that each Floquet eigenstate $U_F|\omega_{\bs,\ell}\rangle = e^{i\omega_{\bs,\ell}}|\omega_{\bs,\ell}\rangle$ can still be labeled by its dominant spin patterns $\pm\bs$ and the associated $\ell$. Namely, $|\omega_{\bs,\ell}\rangle$ can be viewed as a perturbed deformation of the unperturbed cat eigenstate $|\bs,\ell\rangle$. Up to the first order in $\lambda$, the general corrected eigenstate given Eq.~\eqref{eq:deg_eigs} reduces to the specific form as~(see Appendix~\ref{sec3-3} for algebras)
	\begin{align}\nonumber
		|\omega_{\boldsymbol{s},\ell}\rangle &= \frac{1}{\sqrt{N_{\boldsymbol{s}}}} \left[ 
		|\boldsymbol{s}, \ell\rangle 
		+ \sum_{j=1}^L \left(
		X_{\boldsymbol{s},j} |\boldsymbol{s}^{[j]}, \ell\rangle 
		+
		Y_{\boldsymbol{s},j} |\boldsymbol{s}^{[j]}, 1-\ell\rangle
		\right)
		\right],
		\\ \label{eq:eig_corrected_sec2}
		&\eta_{\boldsymbol{s}} =  \sum_{j=1}^L\left( |X_{\boldsymbol{s},j}|^2 + |Y_{\boldsymbol{s},j}|^2 \right), 
		\qquad
		N_{\boldsymbol{s}} = 1 + \eta_{\boldsymbol{s}},
	\end{align}
	where $N_\bs$ is the normalization factor and $\boldsymbol{s}^{[j]}$ is the spin pattern that only differs from $\boldsymbol{s}$ by a spin flip at site $j$, namely,
	\begin{align}
		\boldsymbol{s}^{[j]}: \quad 
		s^{[j]}_{k\neq j} = s_k,\quad 
		s^{[j]}_j = -s_j.
	\end{align}
	The coefficients $X_{\boldsymbol{s},j}$ and $Y_{\boldsymbol{s},j}$ quantify the strength of hybridization between the dominant cat state $|\boldsymbol{s},\ell\rangle$ and other cat states.
	They are determined by the microscopic parameters $(\{J_j\},\{h_j\},\lambda)$ in Eq.~\eqref{eq:UF_general} as
	\begin{align}\nonumber
		X_{\boldsymbol{s},j} &= 
		e^{-i\Delta_{\boldsymbol{s},j}} \frac{\sin\Delta_{\boldsymbol{s},j}}{\lambda \cos h_j}
		\left[
		\sqrt{1+\left(\frac{\lambda \cos h_j}{\sin \Delta_{\boldsymbol{s},j} }\right)^2 }
		-1
		\right],
		\\ \label{eq:xy_sec2}
		Y_{\boldsymbol{s},j} &=  -e^{-i\Delta_{\boldsymbol{s},j}}
		\frac{\cos\Delta_{\boldsymbol{s},j}}{\lambda \sin (h_js_j)} 
		\left[
		\sqrt{1+ \left(\frac{\lambda\sin h_j}{\cos \Delta_{\boldsymbol{s},j} }\right)^2 }
		-1
		\right],
	\end{align}
	where the Ising interaction gap between two spin patterns differing by one spin flip at $j$ reads
	\begin{align}\label{eq:Delta}
		\Delta_{\boldsymbol{s},j} = -s_j(J_{j-1} s_{j-1} + J_j s_{j+1}).
	\end{align}
	We emphasize that this expression has taken into account possible resonance between pairwise eigenstates, i.e., when $|\sin\Delta_{\boldsymbol{s}, j}|,|\cos \Delta_{\boldsymbol{s}, j}| \ll \lambda$, $|X_{\boldsymbol{s},j}|, |Y_{\boldsymbol{s},j}| \rightarrow 1 $ instead of becoming divergent. Meanwhile, as discussed previously, in the non-resonant limit $|\sin\Delta_{\boldsymbol{s}, j}|,|\cos \Delta_{\boldsymbol{s}, j}| \gg \lambda$, $X_{\boldsymbol{s},j}$ and $Y_{\boldsymbol{s},j}$ reduce to the forms derived from non-degenerate perturbation theory,
	\begin{align}\nonumber
		X_{\boldsymbol{s},j} &
		\xrightarrow{\lambda\ll \sin(\Delta_{\boldsymbol{s},j})}
		e^{-i\Delta_{\boldsymbol{s},j}} 
		\frac{\lambda}{2} \frac{\cos (h_js_j)}{\sin\Delta_{\boldsymbol{s},j}}
		+O(\lambda^3),
		\\ \label{eq:non_reso_sec2}
		Y_{\boldsymbol{s},j} &
		\xrightarrow{\lambda \ll \cos(\Delta_{\boldsymbol{s},j})} 
		- e^{-i\Delta_{\boldsymbol{s},j}} \frac{\lambda}{2} \frac{\sin (h_js_j)}{\cos\Delta_{\boldsymbol{s},j}}
		+ O(\lambda^3).
	\end{align}

	Now, physical observables are determined by $X_{\boldsymbol{s},j}, Y_{\boldsymbol{s},j}$ in Eq.~\eqref{eq:xy_sec2} as we show below. The detailed derivations are presented in Appendix~\ref{sec3-4} and \ref{sec3-5}.
	
	\begin{figure*}[t]
		\includegraphics[width=18cm]{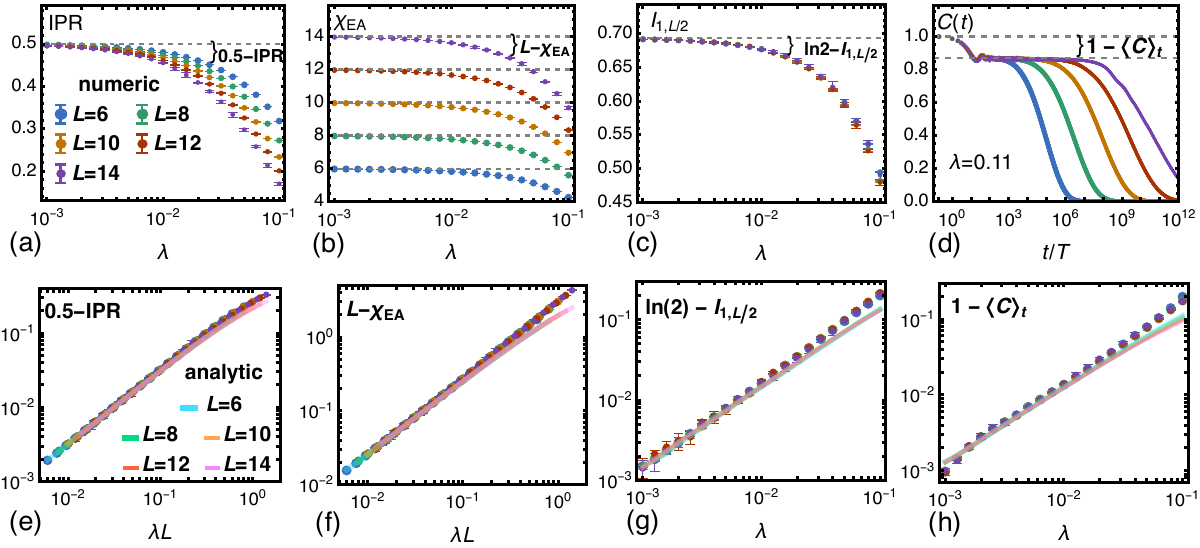}
		\caption{\label{fig1}
			Comparison of numerical results (dots) with analytical predictions in Eqs.~\eqref{eq:ipr}, \eqref{eq:EA_ana}, \eqref{eq:MI_def}--\eqref{eq:S2} and \eqref{eq:corr} (lines) for physical observables in a strongly disordered DTC.
			(a) (e) Inverse participation ratio (IPR) and its deviation from the unperturbed values $0.5$.
			(b) (f) Edwards-Anderson parameter and its deviation from the unperturbed values $L$ for a spin chain of $L$ sites.
			(c) (g) Mutual information for two distant sites $1$ and $L/2$ in a chain with periodic boundary condition, and its deviation from the unperturbed value $\ln(2)\approx0.6931$.
			(d) (h) Evolution of autocorrelator, and the deviation of its steady state plateau from the unperturbed value $1$. $C(t)$ in (d) illustrates an example with $\lambda=0.08$, while the value of $\langle C\rangle_t$ in (h) is extracted by averaging $C(t)$ over the periods $t\in[10^2, 10^3]T$ at each $\lambda$ and $L$. 
			In all cases, $J=1, W=1.6$. Results in (a)--(c) and (e)--(g) are averaged over all eigenstates, while (h) is sampled over all initial Fock states. Error bars, if invisible, are within the size of data points. In all panels except (d), we average over $10^6$, $10^6$, $10^5$, $10^4$, $0.3\sim1.4\times10^4$ samples for $L=6,8,10,12,14$ at each $\lambda$, respectively. In panel (d), we average over $10^5$, $10^5$, $10^4$, $10^4$, $2100$ samples for $L=6,8,10,12,14$.
		}
	\end{figure*}

	{\bf (1)} The inverse participation ratio (IPR) characterizing the Fock space localization of eigenstates is defined as
	\begin{align}
		\text{IPR}(\omega_{\boldsymbol{s},\ell}) = \sum_{\boldsymbol{s}'} \left|
		\langle\boldsymbol{s}'| \omega_{\boldsymbol{s},\ell}\rangle
		\right|^4.
	\end{align}
	Generically, the IPR of an eigenstate approaches $0$ in the fully delocalized limit and assumes larger values in the localized limit. Here, in the unperturbed limit $\lambda=0$, $|\omega_{\boldsymbol{s},\ell}\rangle $ approaches an idealized GHZ state and is equally localized into two Fock states $\left|\pm\bs \right\rangle$, so $\text{IPR} \rightarrow 1/2$. Under perturbation, the eigenstate IPRs take the analytical form
	\begin{align}\label{eq:ipr}
		\text{IPR}(\omega_{\boldsymbol{s},\ell}) = 
		\frac{1}{2N_{\boldsymbol{s}}^2} \left[
		1+\sum_{j=1}^L 
		\left(|X_{\boldsymbol{s},j}|^4 + |Y_{\boldsymbol{s},j}|^4 + 6|X_{\boldsymbol{s},j}|^2|Y_{\boldsymbol{s},j}|^2 \right)
		\right].
	\end{align}
	Note that in the non-degenerate limit $|\sin\Delta_{\boldsymbol{s},j}|, |\cos\Delta_{\boldsymbol{s},j}| \gg \lambda$, we have $|X_{\boldsymbol{s},j}|, |Y_{\boldsymbol{s},j}|\rightarrow 0$ reproducing $\text{IPR}\rightarrow 1/2$.
	
	{\bf (2)} The Edwards-Anderson parameter for spin-glass order of disordered DTCs~\cite{mi2022N,ippoliti2021PQ} is defined as
	\begin{align}\label{eq:EA_def}
		\chi_{\text{EA}}(\omega_{\boldsymbol{s},\ell}) = \frac{1}{L-1} \sum_{j,k; j\neq k}  \langle \omega_{\boldsymbol{s},\ell}| \sigma^z_j \sigma^z_k |\omega_{\boldsymbol{s},\ell}\rangle^2.
	\end{align}
	Note that the double summation $j,k = 1,2,\dots, L$ yields $L(L-1)$ terms under the constraint $j\neq k$. Then, in the unperturbed limit $\lambda=0$ with $\langle\omega_{\boldsymbol{s},\ell} | \sigma^z_j \sigma^z_k|\omega_{\boldsymbol{s},\ell}\rangle^2 = 1$, $\chi_{\text{EA}}\rightarrow L$, which is an extensive quantity. The analytical form under perturbation can be obtained as	
	\begin{align}\label{eq:EA_ana}
		\chi_{\text{EA}} (\omega_{\boldsymbol{s},\ell}) &= 
		L - \frac{8}{N_{\boldsymbol{s}}} \left[
		\eta_{\boldsymbol{s}} - \frac{\eta_{\boldsymbol{s}}^2 + (L-2) \sum_{j=1}^L \left(
			|X_{\boldsymbol{s},j}|^2 + |Y_{\boldsymbol{s},j}|^2
			\right)^2 }{(L-1)N_{\boldsymbol{s}}} 
		\right].
	\end{align}

	{\bf (3)} As a complementary definition of the spontaneous breaking of time translation symmetry for DTCs in terms of eigenstate long-range entanglement~\cite{else2016PRLb}, the mutual information between two distant sites (recall that we take periodic boundary condition) $1$ and $L/2$ is defined as
	\begin{align}\label{eq:MI_def}
		I_{1,L/2}(\omega_{\boldsymbol{s},\ell}) = S_1(\omega_{\boldsymbol{s},\ell}) + S_{L/2}(\omega_{\boldsymbol{s},\ell}) - S_{1,L/2}(\omega_{\boldsymbol{s},\ell}).
	\end{align}
	Here, the entanglement entropy $S_A = -\text{Tr}\left(\rho_A \ln\rho_A\right)$ for the subsystem $A$ is defined via the reduced density matrix $\rho_A = \text{tr}_{j\notin A} \rho$, where sites outside the subsystem $A$ are traced off from the full density matrix $\rho=|\omega_{\boldsymbol{s},\ell}\rangle \langle\omega_{\boldsymbol{s},\ell}|$.
	For an idealized GHZ eigenstate, the mutual information between any two sites are constantly $\ln(2)$ regardless of their separations, while such a value decays to zero in the delocalized limit. With weak perturbations, the mutual information can be similarly calculated perturbatively. First, the explicit form for single-site entanglement entropy reads
	\begin{align}\nonumber
		S_j &= -p^+_j \ln p^+_j - p^-_j\ln p^-_j, \\ \nonumber
		p^\pm_j &=
		\frac{1}{2}\pm \frac{1}{N_{\boldsymbol{s}}}
		\Big[ 
		\left(\text{Re} X_{{\boldsymbol{s}},j} \right)^2 + \left(\text{Im} Y_{{\boldsymbol{s}},j} \right)^2 
		\\ \label{eq:S1}
		& \qquad+
		\left( 
		-\text{Re}\left(X_{\boldsymbol{s},j}^* Y_{\boldsymbol{s},j}\right)
		+
		\sum_{k\neq j} \text{Re}\left(X_{\boldsymbol{s},k}^* Y_{\boldsymbol{s},k}
		\right)
		\right)^2 \Big]^{1/2}.
	\end{align}
	Further, the two-site entanglement entropy can be obtained from the corresponding reduced density matrix
	\begin{align} \nonumber
		&S_{1,L/2} = -\text{Tr}\left(\rho_{1,L/2} \ln \rho_{1,L/2} \right),
		\\ \nonumber
		\rho_{j,k} &= \frac{1}{2N_{\boldsymbol{s}}}
		\begin{pmatrix}
			\alpha_{jk+} & \beta_{j+}^* & \beta_{k+}^* & 0\\
			\beta_{j+} 
			& |\beta_{j+}|^2 + |\beta_{k-}|^2 
			& \beta_{j+}\beta_{k+}^* + \beta_{j-}^* \beta_{k-} & \beta_{k-} \\
			\beta_{k+} 
			& \beta_{j+}^* \beta_{k+} + \beta_{j-} \beta_{k-}^* 
			& |\beta_{j-}|^2 + |\beta_{k+}|^2 
			& \beta_{j-}\\
			0 & \beta_{k-}^* & \beta_{j-}^* & \alpha_{jk-}
		\end{pmatrix},
		\\ \label{eq:S2}
		& \beta_{j\pm} = e^{\pm ih_js_j}\left( X_{\boldsymbol{s},j} \pm Y_{\boldsymbol{s},j}\right),\quad 
		\alpha_{jk\pm} = 1+\sum_{m\neq j,k} |\beta_{m\pm}|^2.
	\end{align}
	Feeding Eqs.~\eqref{eq:S1} for $S_1, S_{L/2}$ and \eqref{eq:S2} for $S_{1,L/2}$ back to Eq.~\eqref{eq:MI_def}, we have the mutual information.

	{\bf (4)} On top of the eigenstate properties shown above, it is also of interest to directly focus on the spatiotemporal dynamics by defining the autocorrelator
	\begin{align}\label{eq:corr}
		C(t) = \frac{(-1)^{t/T}}{L}\sum_{j=1}^L \langle \sigma^z_j(t) \sigma^z_j(0)\rangle_{\boldsymbol{s}},
	\end{align}
	where $\sigma^z_j(t) = (U_F)^{t/T} \sigma^z_j U_F^{t/T}$, and we average over various initial Fock states $|\boldsymbol{s}\rangle$. In a steady state of DTC dynamics, $C(t)$ shows a plateau $\langle C\rangle_t$ representing rigid period-doubled oscillation of local magnetic order $\langle \sigma^z_j(t)\rangle \propto (-1)^{t/T} \langle C\rangle_t$, with the analytical result
	\begin{align}\label{eq:autocorr}
		\langle C\rangle_t &= 1 -
		\left\langle 
		\frac{4}{N_{\boldsymbol{s}}L}
		\left[
		\eta_{\boldsymbol{s}}
		-
		\frac{1}{N_{\boldsymbol{s}}} 
		\sum_{j=1}^L
		\left(|X_{\boldsymbol{s},j}|^2 + |Y_{\boldsymbol{s},j}|^2 \right)^2
		\right]\right\rangle_{\boldsymbol{s}}.
	\end{align}
	In practical calculations, it is more efficient to sample over different random spin patterns $\boldsymbol{s}$ for $\langle\dots \rangle_{\boldsymbol{s}}$, which is the same as averaging over all spin patterns $\frac{1}{2^L}\sum_{\boldsymbol{s}}$.

	\begin{figure}[t]
		\includegraphics[width=8.8cm]{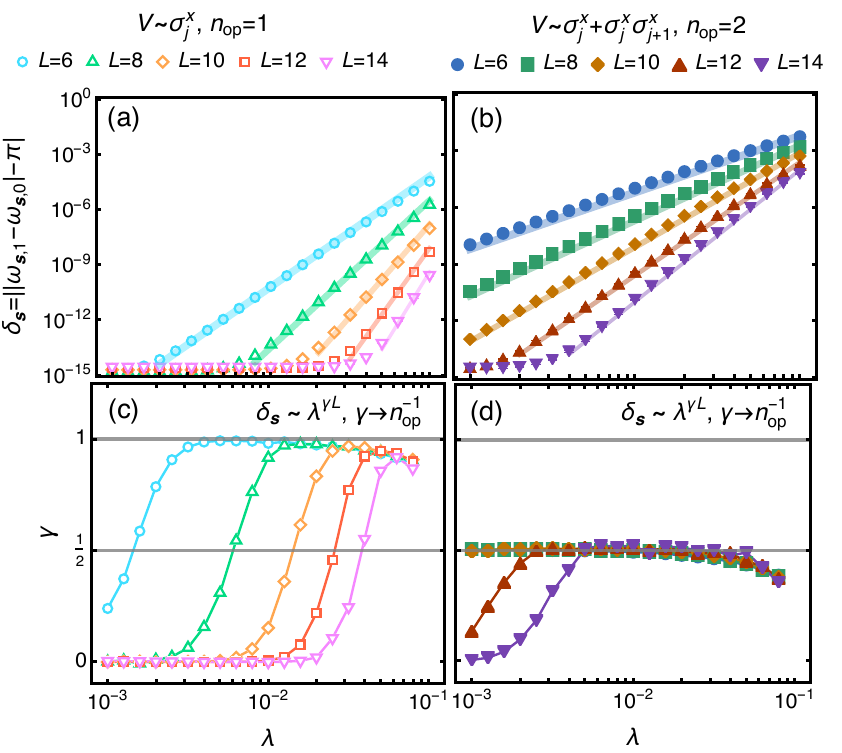}
		\caption{\label{fig2} Scaling of spectral pairing deviation with different operator product order $n_{\text{op}}$ for the perturbation. Here we extract  $\omega_{\boldsymbol{s},0}- \omega_{\boldsymbol{s},1}$ by checking the quasienergy difference between the $m$-th and $(m+2^{L-1})$-th eigenstate, where quasienergies obtained numerically are arranged as $\omega_m < \omega_{m+1}$.
			(a) Single-spin perturbation with $n_{\text{op}}=1$. 
			(b) Single- and two-spin perturbation with $n_{\text{op}}=2$. 
			(c) (d) Extraction of scaling exponent for (a) and (b), respectively, showing that the deviation $\delta_{\boldsymbol{s}} \sim \lambda^{\gamma L}$ from a perfect spectral pairing $|\omega_{\boldsymbol{s},1} - \omega_{\boldsymbol{s},0}|=\pi$ scales as $\gamma \rightarrow n_{\text{op}}^{-1}$ in the strong disorder (weak perturbation) limit, consistent with Eq.~\eqref{eq:sp}. Here for each numerical data point, we averaged over all eigenstates of different dominant spin patterns $\boldsymbol{s}$ and disorder samples. We average over $10^6$, $10^6$, $10^5$, $10^4$ samples for $L=6,8,10,12$, while for $L=14$, we average over $0.3\sim1.4\times10^4$ samples for $n_\mathrm{op}=1$ and $2000$ samples for $n_\mathrm{op}=2$.  In all panels, machine number precision is limited to $\sim 10^{-15}$.
		}
	\end{figure}
	
	{\bf (5)} The lifetime $\tau_*$ of DTC oscillations is related to the spectral gap between pairwise cat eigenstates of the same spin patterns, namely, $\tau_* \propto \delta_{\boldsymbol{s}}^{-1}$. The deviation $\delta_{\boldsymbol{s}}$ scales as
	\begin{align}\label{eq:sp}
		\delta_{\boldsymbol{s}} = \left| |\omega_{\boldsymbol{s},1} - \omega_{\boldsymbol{s},0}| - \pi\right| = O(\lambda^{L/n_{\text{op}}}),
	\end{align}
	where $n_{\text{op}}$ is the operator product order that counts how many spin operators are multiplied in the perturbation terms, i.e., 
	\begin{align}\nonumber
		n_{\text{op}}=1:& \quad V = \sum_{\mu,j} h_j^\mu \sigma^\mu_j, 
		\\ \nonumber
		n_{\text{op}}=2: & \quad 
		V=\sum_{\mu,j} h_j^\mu \sigma^\mu_j
		+ 
		\sum_{\mu\nu,jk} K^{\mu\nu}_{jk} \sigma^\mu_j \sigma^\nu_k,
		\\ \label{eq:nop12}
		& \mu,\nu=x,y,z,\quad j,k = 1,2,\dots, L.
	\end{align}
	Unlike previous properties attributable to the corrected eigenstate, Eq.~\eqref{eq:sp} is obtained by the correction to quasienergies $e^{i\omega_{\bs,\ell}}$ in Eq.~\eqref{eq:qe_correction}. The resolvent formalism enables us to explicitly track the perturbative structure of $\omega_{\bs,0}$ and $\omega_{\bs,1}$, revealing that the two remain identical up to order $L/n_{\mathrm{op}}-1$. The detailed derivation from Eq.~\eqref{eq:qe_correction} to Eq.~\eqref{eq:sp} is a lengthy one, which is given in Appendix~\ref{sec3-4}. But the result in Eq.~\eqref{eq:sp} is easy to use and applicable to a wide range of models, as we shall see later.
	
	It is worth stressing the valid parameter regime to apply the above results. For instance, in Eq.~\eqref{eq:sp}, although the exponential growth of DTC lifetime $\tau_*\propto \delta_{\boldsymbol{s}}^{-1} \sim (1/\lambda)^{L/n_{\text{op}}}$ with system size $L$ may tentatively suggest an infinite lifetime at thermodynamic limit $L\rightarrow\infty$, a prerequisite to apply Eq.~\eqref{eq:sp} is that the pairwise resonance assumption remain valid, and the corrected eigenstate is still dominated by a major Fock state.
	That means the correction terms in Eq.~\eqref{eq:eig_corrected_sec2} must have smaller amplitudes than the dominant part $|\boldsymbol{s},\ell\rangle$.
	Since the resonance probability for the correction parameter $X_{\boldsymbol{s},\ell}, Y_{\boldsymbol{s},\ell}$ to approach 1 is $\propto \lambda$ (due to the pairwise resonance postulation discussed above Eq.~\eqref{eq:mix_matrix}), and there are totally $L$ terms in the summation $\sum_{j=1}^L$ of Eq.~\eqref{eq:eig_corrected_sec2}, we can estimate the amplitudes of correction terms $\propto (\lambda L)$.
	Thus, the valid parameter regime to apply Eq.~\eqref{eq:sp} --- and essentially all the results in this section --- can be estimated as $\lambda L \lesssim 1$. As we shall see in the comparison with numerical results in the next section, i.e. in Figs.~\ref{fig1}(e)--(h), this is indeed the case. 
	
	Analytical results obtained above can be directly used to calculate the values of physical observables with parameters in the model. 
	For {\bf (1)} -- {\bf (4)} concerning eigenstate corrections, one could feed in the parameters $(\{J_j\}, \{h_j\}, \lambda)$ to calculate the key parameters $(X_{\boldsymbol{s},j}, Y_{\boldsymbol{s},j})$ in Eq.~\eqref{eq:xy_sec2} for {\em each} disorder sample and spin patterns $\boldsymbol{s}$, which would then prescribe the physical observables including Eq.~\eqref{eq:ipr} for IPR, Eq.~\eqref{eq:EA_ana} for Edwards-Anderson parameter, Eqs.~\eqref{eq:MI_def}--\eqref{eq:S2} for mutual information, and Eq.~\eqref{eq:autocorr} for the plateau height of autocorrelators in the steady state. Also, quasienergy corrections leads to an exponentially small deviation $\delta_{\boldsymbol{s}}$ in {\bf (5)} for the spectral pairing gap $|\omega_{\boldsymbol{s},1} - \omega_{\boldsymbol{s},0}|$ from the rigid value $\pi$, which can be determined by simply checking the form of perturbation terms.
	
	\subsection{Numerical verification}
	
	We would illustrate the analytical predictions by comparing them with numerical simulation via exact diagonalization. The IPR averaged over all eigenstates is shown in Fig.~\ref{fig1}(a), where we observe a size-dependent deviation from the unperturbed value $\text{IPR}\xrightarrow{\lambda\rightarrow0}0.5$. In Fig.~\ref{fig1}(e), data collapse confirms that the deviation $0.5-\text{IPR}$ scales as a function of $O(\lambda L)$, where numerical data satisfactorily coincides with analytical predictions by Eq.~\eqref{eq:ipr}. The size dependence can be intuitively understood from a perturbative perspective. Starting from the idealized case with perfect GHZ eigenstates, a single-spin perturbation in Eq.~\eqref{eq:UF_nop1} can flip a spin at site $j=1,2,\dots,L$ by $e^{i\lambda V} = 1+i\lambda\sum_{j=1}^L \sigma^x_j + O(\lambda^2)$. Thus, the eigenstate would have a total probability $\propto L$ to be driven out of the original Fock subspace hosting the idealized GHZ state. 
	
	Remaining physical observables related to eigenstate corrections can be analyzed similarly. In Fig.~\ref{fig1}(b), the Edwards-Anderson parameter approaches the system size $L$ in the unperturbed limit $\lambda\rightarrow 0$, while the deviation, as shown in Fig.~\ref{fig1}(f), also scales as $\lambda L$. Nevertheless, the dependence on $L$ here is simply due to that $\chi_{\text{EA}}$ is an extensive quantity, while $\chi_{\text{EA}}$ itself is a two-body observable. This is in contrast to $\text{IPR}$ that is an $L$-body intensive observable. The distinction can be further highlighted by checking the mutual information in Fig.~\ref{fig1}(c). Here, the deviation of $I_{1,L/2}$ from the unperturbed value $\ln(2)$, as shown in Fig.~\ref{fig1}(g), scales as $\lambda$ because $I_{1, L/2}$ is a two-body intensive quantity. Finally, the dynamics of autocorrelator $C(t)$ is exemplified in Fig.~\ref{fig1}(d). We see that after the initial transient period with fast decay, $C(t)$ reaches a plateau representing the steady state with rigid oscillations of local magnetic order. Deviations of the plateau height $\langle C\rangle_t$ from the unperturbed value $1$ is illustrated in Fig.~\ref{fig1}(h), where data collapse confirms the scaling linearly proportional to $\lambda$. In all cases for Fig.~\ref{fig1}(f)--(h), we find quantitative agreements deep inside the strongly disordered regime $J_0, W\gg \lambda$ between the numerical data and the analytical predictions in Eq.~\eqref{eq:EA_ana}, Eqs.~\eqref{eq:MI_def}--\eqref{eq:S2}, and Eq.~\eqref{eq:corr}, respectively.
	
	The lifetime for autocorrelators $C(t)$ to stay in the plateau extends exponentially with the increase of system size, as we see in Fig.~\ref{fig1}(d). This is quantitatively 
	prescribed in our analytical theory that quasienergy corrections for eigenstates dominated by the same spin patterns are identical up to $\lambda^{L-1}$ order (i.e. for single-spin perturbation in Eq.~\eqref{eq:UF_nop1} with $n_{\text{op}}=1$), which leads to the result in Eq.~\eqref{eq:sp}. Then, aside from fluctuations that average to zero in the steady state, the local magnetic order oscillates as
	\begin{align}
		\langle \boldsymbol{s}| \sigma^z_j(t)|\boldsymbol{s}\rangle \propto \cos\left(
		(\pi + \delta_{\boldsymbol{s}}) \frac{t}{T}
		\right)
		=(-1)^{t/T} \cos\left(\delta_{\boldsymbol{s}} \frac{t}{T}\right),
	\end{align}
	which exhibits an envelope $\cos(\delta_{\boldsymbol{s}}t/T)$ for the period-$2T$ oscillations $(-1)^{t/T}$, and implies a lifetime $\tau_* = \pi T/2\delta_{\boldsymbol{s}} \propto \delta_{\boldsymbol{s}}^{-1}$. Thus, knowing the scaling of spectral pairing deviation $\delta_{\boldsymbol{s}}$ equivalently tells the DTC lifetime scaling.
	
	In Fig.~\ref{fig2}, we test the analytical prediction of Eq.~\eqref{eq:sp} for quasienergy corrections against numerical results. For genericness, we adopt two models of different perturbations, with one given in Eq.~\eqref{eq:UF_nop1} representing $n_{\text{op}}=1$ in Eq.~\eqref{eq:nop12}, and the other
	\begin{align}\label{eq:UF_nop2}
		V=\sum_{j=1}^L \left(
		\sigma^x_j + \sigma^x_j \sigma^x_{j+1}
		\right)
	\end{align}
	standing for $n_{\text{op}}=2$. The unperturbed part $U_0$ are both given by Eq.~\eqref{eq:UF_general}. The deviations $\delta_{\boldsymbol{s}}$ from spectral $\pi$-pairing for these two cases are plotted in Fig.~\ref{fig2}(a) and Fig.~\ref{fig2}(b) respectively. Due to the limited machine number precision, the deviation is only resolved for $\delta_{\boldsymbol{s}} \gtrsim 10^{-15}$, within which see the exponential scaling in both cases as discussed in Ref.~\cite{vonkeyserlingk2016PRBa}, and the $n_{\text{op}}=1$ case in Fig.~\ref{fig2}(a) with single-spin perturbation shows a steeper slope compared with the $n_{\text{op}}=2$ case in Fig.~\ref{fig2}(b). Further, to quantify the scaling
	\begin{align}
		\delta_{\boldsymbol{s}} \propto \lambda^{\gamma L},
	\end{align}
	where the exponent $ \gamma L$ represents the slope in Fig.~\ref{fig2}(a) and Fig.~\ref{fig2}(b), we compute
	\begin{align}
		\gamma = \frac{1}{L} \frac{\partial \left(\ln\delta_{\boldsymbol{s}}\right)}{\partial \left(\ln \lambda\right)},
	\end{align}
	and demonstrate the results in Fig.~\ref{fig2}(c) and Fig.~\ref{fig2}(d).
	Within the range where $\delta_{\boldsymbol{s}}$ can be resolved by numerical precision, we see a clear quantization of the exponent $1/\gamma $ towards $n_{\text{op}}$. Specifically, for the single-spin perturbation $V\sim \sigma^x_j$ in Eq.~\eqref{eq:UF_nop1} with $n_{\text{op}}=1$, as denoted by empty symbols in Fig.~\ref{fig2}(c), the exponent $\gamma$ converges $ 1$, while for $V\sim \sigma^x_j + \sigma^x_j\sigma^x_{j+1}$ with $n_{\text{op}}=2$ in Eq.~\eqref{eq:UF_nop2}, as denoted by filled symbols in Fig.~\ref{fig2}(d), $\gamma$ approaches $1/2$ instead. These numerical results, again, are consistent with the analytical prediction of Eq.~\eqref{eq:sp} deep in the strongly disordered regime $J_0, W\gg\lambda$.
	
	\section{More generic scenarios}\label{sec4}
	
	The previous framework is not limited to a specific model, but can, in principle, be generalized to calculate the physical observables for models of the form in Eq.~\eqref{eq:UF_general} with generic perturbation. For illustrations, we would consider several additional examples in this section, which also serve as a summary of procedures for future works to exploit our framework. Further, we would compare the disordered DTCs with a clean one showing long-range entangled cat scars~\cite{huang2023PRBa,bao2024NCa}, and point out their different characters in resonance.
	
	\subsection{Applications to other models}

	Let us first consider generalizing our previous discussions for the uniform single-axis perturbation $V=\sum_{j=1}^L \sigma^x_j$ to a generic single-spin perturbation,
	\begin{align}
		e^{i\lambda V}, \quad V = \sum_{j=1}^L (\theta^x_j\sigma^x_j + \theta^y_j\sigma^y_j + \theta^z_j\sigma^z_j).
	\end{align}
	Without loss of generality, we set 
	\begin{align}
		(\theta^x_j)^2 + (\theta^y_j)^2 = 1,
	\end{align}
	so the spin-flip perturbation strength is absorbed into the parameter $\lambda$. This model can actually be quantified without detailed calculations. To see it, we can first perform a gauge transformation to merge the perturbative longitudinal field with the dominant one
	\begin{align}\nonumber
		&U_F = U_0 e^{i\lambda V} \rightarrow 
		e^{i\lambda\sum_{j=1}^L \theta_j\sigma^z_j} U_F e^{-i\lambda\sum_{j=1}^L \theta_j \sigma^z_j }
		\\ \nonumber
		&= e^{-i\sum_{j=1}^L J_j \sigma^z_j \sigma^z_{j+1} + (h_j - \lambda\theta^z_j) \sigma^z_j} 
		\prod_{j=1}^L \left(-i\sigma^x_j 
		e^{i\lambda (\theta^x_j\sigma^x_j + \theta^y_j\sigma^y_j + \theta^z_j\sigma^z_j) } 
		e^{-i\lambda\theta^z_j \sigma^z_j}
		\right) 
		\\
		&=
		e^{-i\sum_{j=1}^L J_j \sigma^z_j \sigma^z_{j+1} +  (h_j - \lambda\theta^z_j) \sigma^z_j} 
		\prod_{j=1}^L \left(-i\sigma^x_j 
		\right) 
		e^{i\lambda (\theta^x_j\sigma^x_j + \theta^y_j\sigma^y_j) + O(\lambda^2) } 
		.
	\end{align}
	Further, we perform an additional gauge transformation with $\phi_j = \arctan(\theta^y_j/\theta^x_j)$, to rotate the perturbation angles back to the $x$-axis
	\begin{align}\nonumber
		&
		\tilde{U}_F \equiv 
		e^{i \sum_{j=1}^L \phi_j\sigma^z_j/2}
		\left(e^{i\lambda\sum_{j=1}^L \theta_j\sigma^z_j} U_F e^{-i\lambda\sum_{j=1}^L \theta_j \sigma^z_j }\right)
		e^{-i \sum_{j=1}^L \phi_j\sigma^z_j/2}
		\\ \label{eq:modified_model_1}
		&=
		e^{-i\sum_{j=1}^L J_j \sigma^z_j \sigma^z_{j+1} +  (h_j - \lambda\theta^z_j + \phi_j) \sigma^z_j} 
		\prod_{j=1}^L \left(-i\sigma^x_j 
		\right) 
		e^{i\lambda \sum_{j=1}^L \sigma^x_j + O(\lambda^2) } .
	\end{align}
	where we used $ (\prod_{j=1}^L -i\sigma^x_j)e^{-i\phi_j\sigma^z_j/2}
	=
	e^{+i\phi_j\sigma^z_j/2}(\prod_{j=1}^L -i\sigma^x_j)$ in the second step.
	Then, we see that up to the order $\lambda^1$, this model is exactly the same as Eq.~\eqref{eq:UF_general} with a modified longitudinal field $\tilde{h}_j = h_j -\lambda \theta^z_j + \phi_j$. Since all the gauge transformations do not flip spins, previous results characterizing localization properties can be directly applied to Eq.~\eqref{eq:modified_model_1}. 
	
\begin{figure}[H]
		\includegraphics[width=8.5cm]{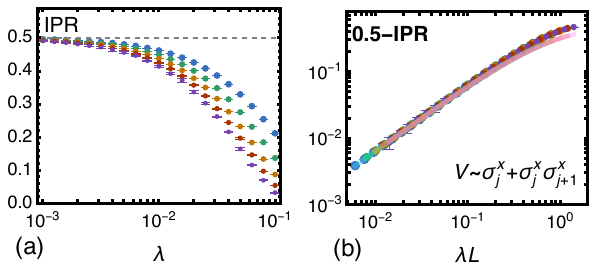}
		\caption{\label{fig3} IPR for the generalized model with both single- and two-spin perturbations in Eq.~\eqref{eq:UF_nop2}.  (a) IPRs averaged over eigenstates and disorder realizations. (b) Deviation of IPRs from the unperturbed value $0.5$, for the model. Legends are the same as in Fig.~\ref{fig1} with colors denoting different system sizes. The analytical results, denoted by lines in panel (b), corresponds to Eq.~\eqref{eq:ipr_n2_sec3}. For $L=6,8,10,12,14$, we sample over $10^6, 10^6, 10^5, 10^4, 2\times 10^3$ disorder realizations at each $\lambda$, respectively. 
		}
	\end{figure}

	We further illustrate the application of our framework with a model involving both the single- and two-spin perturbations	in Eq.~\eqref{eq:UF_nop2},
	\begin{align}\nonumber
		e^{i\lambda V},\qquad
		V = \sum_{j=1}^L (\sigma^x_j + \sigma^x_{j}\sigma^x_{j+1}).
	\end{align}
	The quasienergy corrections for this model have already been examined in Fig.~\ref{fig2}(b) and Fig.~\ref{fig2}(d). Here we
	present the corresponding corrections to the eigenstates. The newly introduced two-spin term $\sigma^x_j\sigma^x_{j+1}$ induces additional hybridization between cat states that differ by flipping two spins simultaneously. As a result, compared with Eq.~\eqref{eq:eig_corrected_sec2}, the first-order corrected eigenstate now contains extra contributions (see Appendix~\ref{sec-2spin} for algebraic details),
	\begin{align}\nonumber
		|\omega_{\boldsymbol{s},\ell}\rangle
		&=
		\frac{1}{\sqrt{N_{\boldsymbol{s}}}}
		\Bigg[
		|\boldsymbol{s},\ell\rangle
		+
		\sum_{j=1}^L
		\left(
		X_{\boldsymbol{s},j}|\boldsymbol{s}^{[j]},\ell\rangle
		+
		Y_{\boldsymbol{s},j}|\boldsymbol{s}^{[j]},1-\ell\rangle
		\right)
		\\ \nonumber
		&\qquad
		+
		\sum_{j=1}^L
		\left(
		X'_{\boldsymbol{s},j,j+1}|\boldsymbol{s}^{[j,j+1]},\ell\rangle
		+
		Y'_{\boldsymbol{s},j,j+1}|\boldsymbol{s}^{[j,j+1]},1-\ell\rangle
		\right)
		\Bigg],
		\\ \nonumber
		N_{\boldsymbol{s}} &= 1+\eta_{\boldsymbol{s}},
		\\ \label{eq:eig_corrected_2spin}
		\eta_{\boldsymbol{s}}
		&=
		\sum_{j=1}^L
		\left(
		|X_{\boldsymbol{s},j}|^2
		+
		|Y_{\boldsymbol{s},j}|^2
		+
		|X'_{\boldsymbol{s},j,j+1}|^2
		+
		|Y'_{\boldsymbol{s},j,j+1}|^2
		\right).
	\end{align}
	where $\boldsymbol{s}^{[j,j+1]}$ denotes the spin configuration obtained from $\boldsymbol{s}$ by flipping the spins at sites $j$ and $j+1$,
	\begin{align}\nonumber
		\boldsymbol{s}^{[j,j+1]}:\quad
		s^{[j,j+1]}_{k\neq j,j+1}=s_k,\quad
		s^{[j,j+1]}_j=-s_j,\quad
		s^{[j,j+1]}_{j+1}=-s_{j+1},
	\end{align}
	and the corresponding Ising interaction gap is
	\begin{align}
		\Delta_{\boldsymbol{s},j,j+1}
		=
		\frac{E_{\boldsymbol{s},\ell}-E_{\boldsymbol{s}^{[j,j+1]},\ell}}{2}
		=
		-J_{j-1}s_{j-1}s_j
		-
		J_{j+1}s_{j+1}s_{j+2},
	\end{align}
	with $E_{\boldsymbol{s},\ell}=
	-\sum_{j=1}^L J_js_js_{j+1}
	+
	\pi\ell
	$
	given by Eq.~\eqref{eq:U0_solutions_sec2}.
	Parameters $X_{\boldsymbol{s},j}, Y_{\boldsymbol{s},j}$ are the same as in Eq.~\eqref{eq:xy_sec2}, while
	new hybridization coefficients $X'_{\boldsymbol{s},j,j+1}$ and $Y'_{\boldsymbol{s},j,j+1}$ are
	\begin{align}\nonumber
		X'_{\boldsymbol{s},j,j+1}
		&=
		e^{-i\Delta_{\boldsymbol{s},j,j+1}}
		\frac{\sin\Delta_{\boldsymbol{s},j,j+1}}
		{\lambda\cos(h_js_j+h_{j+1}s_{j+1})}
		\\ \nonumber
		&\qquad
		\cdot
		\left[
		\sqrt{
			1+
			\left(
			\frac{\lambda\cos(h_js_j+h_{j+1}s_{j+1})}
			{\sin\Delta_{\boldsymbol{s},j,j+1}}
			\right)^2
		}
		-1
		\right],
		\\
		Y'_{\boldsymbol{s},j,j+1}
		&=
		-
		e^{-i\Delta_{\boldsymbol{s},j,j+1}}
		\frac{\cos\Delta_{\boldsymbol{s},j,j+1}}
		{\lambda\sin(h_js_j+h_{j+1}s_{j+1})}
		\\ \nonumber
		&\qquad
		\cdot
		\left[
		\sqrt{
			1+
			\left(
			\frac{\lambda\sin(h_js_j+h_{j+1}s_{j+1})}
			{\cos\Delta_{\boldsymbol{s},j,j+1}}
			\right)^2
		}
		-1
		\right].
	\end{align}
	These coefficients incorporate the effect of pairwise eigenstate resonances similar to the case with single-spin perturbation.
	To test the results, we calculate the IPR of corrected eigenstate (see Appendix~\ref{sec-2spin} for details),
	\begin{align}\nonumber
		&\text{IPR}(\omega_{\boldsymbol{s},\ell}) 
		= 
		\sum_{\boldsymbol{s}'} \left|
		\langle \boldsymbol{s}'| \omega_{\boldsymbol{s},\ell} \rangle
		\right|^4
		\\  \nonumber
		&=
		\frac{1}{2N_{\boldsymbol{s}}^2}
		\left[
		1
		+
		\sum_{j=1}^L 
		\left(
		|X_{\boldsymbol{s},j}|^4 + |Y_{\boldsymbol{s},j}|^4 
		+
		6|X_{\boldsymbol{s},j}|^2|Y_{\boldsymbol{s},j}|^2
		\right)
		\right.
		\\ \label{eq:ipr_n2_sec3}
		&\quad 
		\left.
		+\sum_{j=1}^L \left( 
		|X'_{\boldsymbol{s},j,j+1}|^4 + |Y'_{\boldsymbol{s},j,j+1}|^4
		+
		6|X'_{\boldsymbol{s},j,j+1}|^2 |Y'_{\boldsymbol{s},j,j+1}|^2
		\right) 
		\right].
	\end{align}
	We compare the analytical prediction to numerical data in Fig.~\ref{fig3}. Here, we similarly observe a size-dependent deviation of IPR from the unperturbed limit $1/2$ in Fig.~\ref{fig3}(a), like in Fig.~\ref{fig1}(a). The deviation $1/2-\text{IPR}$, as shown in Fig.~\ref{fig3}(b), agree with the analytical results in Eq.~\eqref{eq:ipr_n2_sec3}. 
	
	\subsection{Comparison with cat scar DTCs in the clean limit}
	
	In the previous analysis, we have fixed the strong disorders $J_j, h_j$, and focused on the effects of different perturbation strengths $\lambda$.
	Beyond it, analytical framework introduced in this work also offers a unified description for DTCs with different disorder strength. In particular, in the clean limit, majority eigenstates exhibit Fock space prethermal behaviors~\cite{Bao2026}, while four rare cat scar eigenstates with ferromagnetic (FM) and anti-ferromagnetic (AFM) patterns possess long-range entanglement~\cite{huang2023PRBa,bao2024NCa}, similar to the cat eigenstates in the strongly disordered cases discussed in this work. We shall compare the analytical perturbation quantification for these cat scars with the majority cat eigenstates in the disordered scenarios. We would see that the cat scars are free from pairwise resonances and therefore show different scaling behaviors for physical observables.

	With translation invariance, site dependence vanishes, and we have $h_j\rightarrow h$, $\Delta_{\boldsymbol{s},j} = -Js_j(s_{j+1} + s_{j-1}) = \pm 2J$, where ``$+$'' for AFM and ``$-$'' for FM patterns, respectively. Explicitly, they take the form in the unperturbed limit as
	\begin{align}\nonumber
		|\text{FM},\ell\rangle &=\frac{1}{\sqrt{2}}\left[ e^{-ihL/2} \left|\uparrow\uparrow\uparrow\uparrow\dots \right\rangle +\left( -1 \right)^\ell
		e^{i hL/2} \left|\downarrow \downarrow\downarrow\downarrow \dots \right\rangle \right],
		\\
		|\text{AFM},\ell\rangle &=\frac{1}{\sqrt{2}}\left[ \left| \uparrow\downarrow\uparrow\downarrow\dots \right\rangle +\left( -1 \right)^\ell
		\left|\downarrow\uparrow\downarrow\uparrow\dots \right\rangle \right],
	\end{align}
	where $\uparrow$ and $\downarrow$ denote $s_j=+1$ and $s_j=-1$, respectively.
	Since the FM and AFM patterns possess extreme number of domain walls, they are non-degenerate with any other eigenstates for the Ising interaction energy~\cite{huang2023PRBa}. Thus, we can exploit the expansion in Eq.~\eqref{eq:non_reso_sec2} to replace
	\begin{align}\nonumber
		X_{\boldsymbol{s},j} &
		\rightarrow 
		|X|=
		\left| \frac{\lambda}{2} \frac{\cos h}{ \sin 2J} \right|
		+O(\lambda^3),
		\\ \label{eq:xy_scar}
		Y_{\boldsymbol{s},j} &
		\rightarrow
		|Y|=
		\left|\frac{\lambda}{2} \frac{\sin h}{ \cos2J} \right|
		+ O(\lambda^3),
	\end{align}
	which is free from disorders and identical for the four scars. We readily see from this expression that the denominator $\sin(2J), \cos(2J)$ for large Ising interaction $J=1$ is never smaller than the perturbation strength $\lambda\ll1$, and therefore resonances are fully suppressed.	Then, Eq.~\eqref{eq:ipr} directly gives the disorder-free result
	\begin{align}\label{eq:IPR_scar}
		\text{IPR}_{\text{scar}} &=
		\frac{1}{2N^2} + O(\lambda^4) =
		\frac{1}{2}- \lambda^2 L \bar{V}_1^2 +O(\lambda^4)
		,
	\end{align}
	where the normalization constant is simplified into
	\begin{align}\nonumber
		&N = 1+ \eta, 
		\quad 
		\eta = \lambda^2L\bar{V}_1^2,
		\\\label{eq:norm_scar}
		& \bar{V}_1^2 = \frac{1}{4}  \left[
		\frac{\cos^2(h)}{ \sin^2(2J) }
		+
		\frac{\sin^2(h)}{ \cos^2(2J)}
		\right].
	\end{align}
	Further, the Edwards-Anderson parameter for spin-glass order of scar eigenstates can be obtained from Eq.~\eqref{eq:EA_ana} with the same substitutions as in Eq.~\eqref{eq:norm_scar},
	\begin{align}\label{eq:EA_scar}
		\chi_{\text{EA}}
		=
		L - \frac{8\eta}{N} +O(\lambda^4)
		= L - 8\lambda^2 L \bar{V}_1^2 +O(\lambda^4).
	\end{align}
	We emphasize again that analytical results for physical observables in Eqs.~\eqref{eq:IPR_scar} -- \eqref{eq:EA_scar} are solely determined by the parameters $(J_j,h_j,\lambda) = (J,h,\lambda)$ in Eqs.~\eqref{eq:UF_general} and \eqref{eq:UF_nop1} without any fitting parameter, for a non-integrable disorder-free model.

	\begin{figure}[h]
		\includegraphics[width=8.5cm]{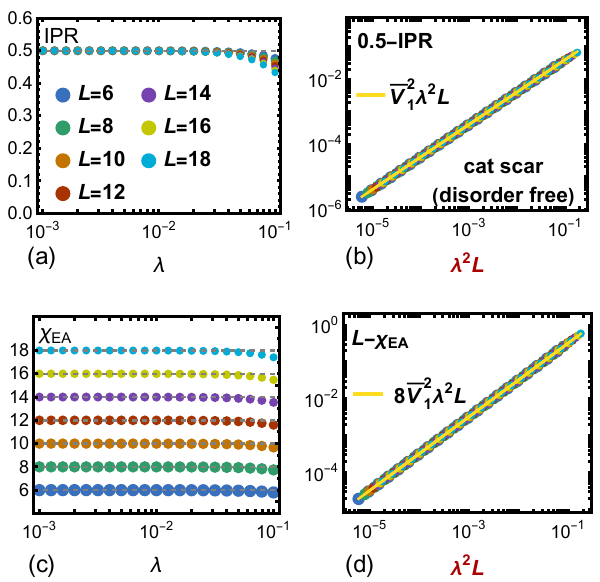}
		\caption{\label{fig4} (a) Eigenstate IPRs and (b) their deviation from the unperturbed value $0.5$ for translation-invariant models hosting rare cat scar eigenstates~\cite{huang2023PRBa}. (c) Edwards-Anderson parameter and (d) their deviation from the unperturbed value $L$ for cat scars. Notably, in panels (b) and (d) for the data collapse, we see that since cat scars are free from resonance, the deviation $0.5-\text{IPR}, L-\chi_{\text{EA}} \sim \lambda^2 L$ is suppressed at the same strength of perturbation $\lambda\ll1$, compared with the majority eigenstates in disordered situation showing stronger deviations $\sim \lambda L$ in Fig.~\ref{fig1}(a), (b), (e), and (f). Here we take $J_j=J=1, h_j=h=0.3$ for the model in Eq.~\eqref{eq:UF_general} and \eqref{eq:UF_nop1}.
		}
	\end{figure}
	We compare the analytical results for cat scars with numerical simulations in Fig.~\ref{fig4}. From Figs.~\ref{fig4}~(a) and (c), we note that at each $\lambda$, IPR and Edwards-Anderson parameters for cat scars are much closer to their respective unperturbed values $0.5$ and $L$, compared with the disordered cases in Fig.~\ref{fig1}(a) and (b). This is further highlighted by the different scaling behaviors for the deviations in terms of $\lambda^2 L$ shown in Figs.~\ref{fig4}~(b) and (d), to be contrasted against Figs.~\ref{fig1}~(e) and (f), respectively, where the deviations of IPR and Edwards-Anderson parameters from unperturbed values scale as $\lambda L$ instead. Thus, with the same strength of perturbation $\lambda$, cat scars in the translation invariant case suffer from less destruction than what the majority of cat eigenstates in disordered cases experience. 
	
	The qualitatively different scaling behaviors originates from that cat eigenstates in disordered cases generally undergo pairwise resonances, while the rare cat scars in translation invariant cases do not. We can see such differences and their consequences from the perturbation theory perspective. Generically, the first-order perturbed eigenstates in both cases take the form $|\omega_{\boldsymbol{s},\ell}\rangle = (1/\sqrt{N}) \left[|\boldsymbol{s},\ell\rangle + \sum_{\boldsymbol{s}'} X_{\boldsymbol{s}'}|\boldsymbol{s}', \ell\rangle + Y_{\boldsymbol{s}'} |\boldsymbol{s}', 1-\ell\rangle \right]$, where $|\boldsymbol{s},\ell\rangle$ denotes an unperturbed cat eigenstate as in Eq.~\eqref{eq:U0_solutions_sec2}, $X_{\boldsymbol{s}'}, Y_{\boldsymbol{s}'}$ are amplitudes of corrections, and $N$ is the normalization constant. If resonance occurs between two unperturbed eigenstates of patterns $\boldsymbol{s}$ and  $\boldsymbol{s}'$, the amplitudes $X_{\boldsymbol{s}'}, Y_{\boldsymbol{s}'} \rightarrow 1$ representing strong hybridization. This will always occur in the strongly disordered scenario because the Ising energy gaps $\Delta_{\boldsymbol{s},j} = -s_j(J_{j-1} s_{j-1} + J_j s_{j+1})$, $s_j=\pm1$, for a random landscape $\{J_j\}$ have nonzero probability to approach 0. More technically, the resonance condition is $\lambda \gtrsim \min(\sin\Delta_{\boldsymbol{s},j}, \cos\Delta_{\boldsymbol{s},j})$, as given by Eq.~\eqref{eq:non_reso_sec2}. Assuming a smooth probability distribution function for $\{\Delta_{\boldsymbol{s},j}\}$, the probability for resonance to occur is proportional to $ \lambda$, giving rise to the leading order corrections that scale as $\lambda$ after averaging over disorders and spin patterns in Fig.~\ref{fig1}(e) -- (h). In contrast, cat scars have large gap $|\Delta|=2J$ away from all other eigenstates in the unperturbed limit. Thus, the perturbed eigenstate takes the form $|\omega_{\boldsymbol{s},\ell}\rangle = (1/\sqrt{N}) \left[ |\boldsymbol{s},\ell\rangle + \lambda \left(  \sum_{\boldsymbol{s}'} C_1|\boldsymbol{s}',\ell\rangle + C_2|\boldsymbol{s},\ell\rangle \right) \right] $, namely, the amplitudes for corrections $X_{\boldsymbol{s}'}, Y_{\boldsymbol{s}'}$ are always of the perturbative strength $\sim \lambda$, an expected result for a generic non-degenerate perturbation theory. Then, the normalization constant becomes $N = 1+ O(\lambda^2)$, such that the unperturbed cat scar $(1/\sqrt{N})|\boldsymbol{s},\ell\rangle$ has a rescaled amplitude $\sim 1-O(\lambda^2)$, which gives rise to the scaling $\sim \lambda^2$ in Fig.~\ref{fig4}(b) and Fig.~\ref{fig4}(d).

	\section{Conclusion and outlook}\label{sec5}
	
	In this work, we have constructed a Floquet resolvent perturbation theory from a generalized Fock space perspective. We obtained closed-form formulae to quantify a set of physical observables essential to the long-range entangled DTCs in the strongly disordered regime, and observe quantitative agreement between analytical predictions and numerical data. To our best knowledge, we are unaware of any previous work that provides a practical way to calculate physical observables analytically in generic disordered DTCs, i.e. for non-integrable models with arbitrary local perturbations without using fitting parameters. Analytical procedures in this work can be applied to generic models with dominant Ising interaction and/or longitudinal fields under local perturbations, and therefore offers an alternative path to quantify DTCs without relying on the debated MBL scheme. This may be particularly useful for the benchmark of experiments in quantum devices with large system sizes inaccessible to numerics. 
	
	We illustrate our results in detail with a concrete example, where we obtain the eigenstate IPR in Eq.~\eqref{eq:ipr}, Edwards-Anderson parameter in Eq.~\eqref{eq:EA_ana}, the mutual information between distant sites in Eq.~\eqref{eq:MI_def}--\eqref{eq:S2}, and the spatiotemporal correlator in Eq.~\eqref{eq:corr}, which are well captured by the first-order corrected eigenstates. Also, the Floquet resolvent formalism conveniently provides analytical forms of corrected quasienergy to arbitrarily higher-orders without needing iterations, with which we derive the quasienergy spectral gap between pairwise cat eigenstates $|\omega_{\boldsymbol{s},1} - \omega_{\boldsymbol{s},0}|$. We find that the deviation of such a gap from the unperturbed value $\pi$ scales as $O(\lambda^{L/n_{\text{op}}})$, where the operator product order $n_{\text{op}}$ means the perturbations involve up to $n_{\text{op}}$-spin terms. Such a scaling is quantitatively verified numerically.
	See also a recent work showing that the full distribution function of spectral pairing deviations can be well fitted by log-normal functions, based on combined analytic and numerical arguments~\cite{penner2025prb}.
	
	More broadly, we have demonstrated the procedures to obtain closed-form formulae for physical observables with additional models, including those with arbitrary single-spin perturbation, and those with both single- and two-spin perturbations. The resonance between pairwise cat eigenstates in disordered cases are highlighted by contrasting to non-resonant cat scars in translation invariant scenarios. We note that resonances in disordered DTCs have led to stronger deviation of physical observables from their unperturbed values as $O(\lambda)$, in contrast to the non-resonant cat scars that exhibit suppressed deviations as $O(\lambda^2)$. 
	
	We should emphasize the logical relations among these perturbatively calculated quantities, especially about their validity as the system size $L$ grows. 
	For any dynamical quantities, the corrected eigenstate will determine their oscillation amplitude, while the spectral pairing for quasienergies will dictate their oscillation lifetime. Thus, to apply the results in our theory, it is necessary for
	Fock space localization to occur. Mathematically, in the proof of spectral pairing, we have assumed that the eigenstates still involve a dominant component made of a GHZ state, which means the eigenstate localization in Fock space is a prerequisite. 
		
	The regime of the parameter $\lambda$ for our theory to be valid is therefore given by the condition $\lambda L\lesssim1$ at $J,W\sim1$ (except for cat scars with a wider validity regime as $\lambda^2L\lesssim1$ at $J\sim1$ ). This is verified in the comparison with numerical results.  One can understand the scaling in the following intuitive way. A spin-flip perturbation would have a probability $\lambda$ to flip a spin and take the eigenstate out of its original unperturbed GHZ subspace. Or, equivalently, there is a probability $(1-\lambda)$ for the eigenstate to remain in its unperturbed form. Since there are $L$ sites, the probability for an entire perturbed eigenstate to stay in its original GHZ subspace is $\propto (1-\lambda)^L \approx 1-\lambda L$. 
	Thus, this work is supposed to quantify systems with a finite and large system sizes $L$, which yields a finite range $\lambda \lesssim 1/L$, but does {\em not} concern systems in the thermodynamic limit $L\rightarrow\infty$. Whether or not MBL-DTC can persist into thermodynamic limit is pending future studies, given current uncertainties of MBL itself as a stable phase of matter.

	It is of interest to compare the parameter regimes in our case with those in the traditional description of MBL. Let us unify notations first. In our work, we fix the disorder strength $W$ while changing the perturbation $\lambda$. Thus, we can view $W/\lambda$ as the ``disorder strength" both here and in the notations of previous works because in those cases, $\lambda$ is fixed for a static Hamiltonian. Our Fock space localized regime corresponds to the parameter range $(\lambda L)\lesssim1$. That would imply a ``critical disorder strength" $W_*/\lambda$ in the language of previous work as $W_*/\lambda \gtrsim L $, which drifts linearly with system size $L$ to infinity as $L$ approaches thermodynamic limit.
	This resembles what was found in a number of numerical works searching for possible MBL transition or crossover~\cite{Suntajs2020,Suntajs2020a,Sels2021,Sierant2020,Sierant2021,DeTomasi2021,Sierant2023,Sels2023}. A linear drift $W_*/\lambda \propto L$ (and in certain cases sublinear) was found in these works, and was cited by some of them as implications for the absence of MBL phase because it diverges in the thermodynamic limit. It is therefore worth future works to study the relations between these drifts found numerically for the LIOM regime, whose underlying mechanisms remain largely unknown, and the drift inherently exist in our case.

	Results in this work open the possibility for several other future investigations. One direction is to formulate a probabilistic description of gaps between every pair of cat eigenstates in the unperturbed limit, which can lead to a more systematic characterization of resonances. Specifically, in this work, our analytical equations have prescribed the conditions for eigenstate resonance to occur. It is desirable to specify, preferably in analytical fashions, which pairs of eigenstates dominate the resonance, and formulate a probabilistic description to quantify such resonances. This may shed light on several intriguing puzzles found in Ref.~\cite{morningstar2022PRBa}. For instance, it is found that deep inside the strongly disordered regime, consecutive gaps do not show level repulsions, but systemwide resonance already proliferates among non-consecutive levels. Finally, the Fock space localization considered in this work serves as a stronger condition for ergodicity breaking than the MBL defined by LIOM or multifractal dimensions. Thus, our quantitative description provides a conservative estimation of the necessary disorder strength for a given number of qubits to exhibit localization, which can be applied to predict new phenomena therein without relying on the assumptions of other MBL theories.
	

	\section*{Acknowledgements}
	This work is supported by the National Natural Science Foundation of China Grant No. 12174389.

	\appendix

	\section{Analytical framework}\label{sec3}
	
	Analytical predictions for physical observables constitute the major results of our work. In this Appendix, we would present the derivations of these results and a more generic framework for arbitrary perturbations based on the Floquet resolvent formalism we develop.

	\subsection{General Floquet resolvent formalism}\label{sec4-1}
	
	Consider a general Floquet unitary of the form $U = U_0e^{i\lambda V}$, where $U_0 = \sum_\alpha e^{iE_\alpha} |E_\alpha \rangle \langle E_\alpha|$ is the rigorously solvable part, and $U = \sum_\alpha e^{i\omega_\alpha}|\omega_\alpha\rangle \langle\omega_\alpha|$ incorporates the perturbation effects of $V$. Here $E_\alpha$ and $\omega_\alpha$ are the unperturbed and corrected quasienergy, respectively. We define two resolvents in the complex plane $z\in\mathbb{C}$, 
	\begin{align}
		G(z)=\frac{1}{z-U},\quad 
		G_0(z) = \frac{1}{z-U_0}.
	\end{align}
	Note that $(z-U_0)=(z-U)+(U-U_0)$, we have $G=G_0+G_0(U-U_0)G$, which gives the Floquet-Born series,
	\begin{equation}\label{eqM:Born-Floquet}
		G\left( z \right) =\sum_{n=0}^{\infty}{G_0\left( z \right) \left[ U_0\left( e^{i\lambda V}-1 \right) G_0\left( z \right) \right] ^n}.
	\end{equation}
	The astute readers would recognize the similarity between resolvents and Green's functions, while the key difference is the complex-valued variable $z$ that will play essential roles in the resolvent formalism.
	
	In a strongly disordered system without symmetry, it is reasonable to assume that no rigorous degeneracy occurs for eigenstates. Then, we can always choose a contour ${\mathscr C}(\alpha)$ in the complex plane that encloses only one single quasienergy $E_\alpha$ and its corrected value $\omega_\alpha$, which should reside close to each other given small perturbations $\lambda\ll1$. Thus, the contour integral of resolvents gives the eigenstate projectors,
	\begin{align}\nonumber
		P_0(\alpha) &=\frac{1}{2\pi i}\oint_{{\mathscr C}(\alpha)}{G_0}(z)\, \mathrm{d}z = |E_\alpha\rangle \langle E_\alpha|
		\\ 
		\label{eqM:projector}
		P(\alpha) &=\frac{1}{2\pi i}\oint_{{\mathscr C}(\alpha)} G(z) \, \mathrm{d}z = |\omega_\alpha\rangle \langle\omega_\alpha|,
	\end{align}
	where the second steps for both $P$ and $P_0$ are obtained by expanding resolvents in their respective eigenbases, i.e. $G(z) = \sum_n (z-\omega_\alpha)^{-1} |\omega_\alpha \rangle \langle \omega_\alpha|$, and applying the residual theorem,
	\begin{align}
		\oint_{\mathscr{C}(\alpha)} \frac{dz}{2\pi i} \frac{1}{(z-\omega_\alpha)^m} 
		=
		\begin{cases}
			0, & m\neq 1;\\
			1, & m=1.
		\end{cases}
	\end{align}
	Substituting the Floquet-Born series of Eq.~\eqref{eqM:Born-Floquet} into Eq.~\eqref{eqM:projector}, the corrected and bare projectors are related as
	\begin{align}\nonumber
		P(\alpha) &= P_0(\alpha) + \sum_{n=1}^\infty \delta P_n(\alpha),
		\\ \label{eq:deltaP}
		\delta P_n(\alpha) &=  \frac{1}{2\pi i} \oint_{{\mathscr C}(\alpha)} G_0(z) \left[
		U_0 (e^{i\lambda V}-1) G_0(z)
		\right]^n \mathrm{d}z,
	\end{align}
	where we note $\delta P_n(\alpha)$ is at least of order $\left[ \dots (e^{i\lambda V}-1)\dots \right]^n \sim O(\lambda^n)$. Then, the perturbation series for corrected eigenstates can be written as
	\begin{align}\label{eq:eigcorrection}
		|\omega_\alpha \rangle = \frac{1}{\sqrt{N}} \left(
		|E_\alpha\rangle + \sum_{n=1}^\infty \delta P_n(\omega_\alpha) |E_\alpha\rangle 
		\right),
	\end{align}
	where $N$ is a normalization constant.

	Meanwhile, to obtain the corrected quasienergy $e^{i\omega_\alpha}$ on top of the unperturbed $e^{iE_\alpha}$, we note that $UG=zG-1$, and therefore $UP(\alpha) = \frac{1}{2\pi i}\oint_{{\mathscr C}(\alpha)} {UG\left( z \right) \mathrm{d}z} = \frac{1}{2\pi i} \oint_{{\mathscr C}(\alpha)} {zG\left( z \right) \mathrm{d}z}$, where in the last step we used $\oint_{\mathscr{C}(\alpha)} 1\mathrm{d}z = 0$. Subtracting $e^{iE_\alpha}P(\alpha)$ from both sides, we have
	\begin{align}
		&\left( U-e^{iE_\alpha} \right) P(\alpha)
		=
		\frac{1}{2\pi i} \oint_{{\mathscr C}(\alpha)}
		\left( z-e^{iE_\alpha} \right) G(z) \, \mathrm{d}z.
	\end{align}
	Plugging in the Floquet-Born series in Eq.~\eqref{eqM:Born-Floquet}, and taking trace on both sides, we arrive at the quasienergy perturbation series
	\begin{align} \nonumber
		& 
		e^{i\omega_\alpha} - e^{iE_\alpha}   \equiv 
		\sum_{n=1}^\infty \text{Tr} \left[ \delta \Omega_n(\alpha) \right],
		\\ \label{eq:qecorrection}
		& 
		\delta \Omega_n(\alpha) = \frac{1}{2\pi i}
		\oint_{{\mathscr C}(\alpha)}
		\left(z-e^{iE_\alpha}\right) G_0(z) \left[U_0 \left(e^{i\lambda V}-1 \right) G_0(z)\right]^n 
		\mathrm{d}z.
	\end{align}
	
	The series in Eqs.~\eqref{eq:eigcorrection} and \eqref{eq:qecorrection} give the corrections for eigenstates and quasienergy, respectively. In the resolvent method, corrections of arbitrary orders are explicitly given by $\delta P_n$ and $\delta \Omega_n$ without iteration, as we see in Eqs.~\eqref{eq:deltaP} and \eqref{eq:qecorrection}. This is particularly useful in proving the spectral pairing that involves quasienergy corrections up to order $\lambda^L$.

	\subsection{Review of the zeroth order eigenstructure}

	Let us apply the general formation to our studies of disordered DTCs. We would start in this subsection by reviewing the eigenstructures in the unperturbed limit~\cite{else2016PRLb}, which constitutes the zeroth-order terms in the perturbation series. At $\lambda=0$, the eigenstructure of $U_0$ has already been given in Eq.~\eqref{eq:U0_solutions_sec2}, and we rewrite it here for completeness,
	\begin{align}\nonumber
		U_0|\bs,\ell\rangle&=e^{iE_{\bs,\ell}}|\bs,\ell
		\rangle,
		\quad E_{\bs,\ell}=-\sum_{j=1}^L\left( J_js_js_{j+1} \right)+\ell \pi,
		\quad
		\ell = 0,1;
		\\ \label{eq:U0_solutions}
		|\bs,\ell\rangle &=\frac{1}{\sqrt{2}}\left[ e^{-i\sum_{j=1}^L h_js_j/2}|\bs\rangle +\left( -1 \right)^\ell
		e^{i\sum_{j=1}^L h_js_j/2} \left|-\bs \right\rangle \right].
	\end{align}
	The zeroth order eigenstates $|\boldsymbol{s},\ell\rangle$ are Fock space localized onto the two bases $\left|\pm\bs\right\rangle$.  Thus, an initial Fock state $|\bs\rangle$ would simultaneously overlap with pairwise eigenstates $|\bs,0\rangle$ and $|\bs,1\rangle$, whose interference produces a Fock space oscillation $U_F^{t/T}|\bs\rangle \propto \left|(-1)^{t/T}\bs \right\rangle$ with period $2T$. Correspondingly, local observables, i.e. the local magnetic order $\langle\boldsymbol{s}| \sigma^z_j(t) |\boldsymbol{s}\rangle$, exhibit similar oscillations between $\pm1$ with doubled periodicity, as
	\begin{align}
		\sigma_j^z(t) = (U_F^\dagger)^{t/T} \sigma^z_j U_F^{t/T} \propto (-1)^{t/T} \sigma^z_j.
	\end{align}

	Under perturbation, DTC dynamics would receive corrections on two aspects from the underlying eigenstructure. On the one hand, the amplitude of such rigid subharmonic oscillations relies on the localization of eigenstates onto pairwise Fock states $\left|\pm\boldsymbol{s} \right\rangle$, henceforth determined by the corrections to eigenstates $|\boldsymbol{s},\ell\rangle$. Meanwhile, the quasienergy correction to pairwise cat eigenstates would result in an envelope modulation, which determines the DTC lifetime in a finite-size system. In the following, we shall incorporate higher order correction to the perturbation series,
	\begin{align}\nonumber
		&
		\exp\left(i\omega_{\bs,\ell}\right) = \exp\left[i\left(E_{\bs,\ell} + \sum_{n=1}^\infty \lambda^n \omega_{\bs,\ell}^{(n)} \right) \right] ,
		\\ 
		&
		|\omega_{\bs,\ell}\rangle 
		=
		\frac{1}{\sqrt{N_{\boldsymbol{s}}}} \left(
		|\bs,\ell\rangle + \sum_{n=1}^\infty \frac{(i\lambda)^n}{n!} |\omega_{\bs,\ell}^{(n)}\rangle
		\right),
	\end{align}
	where $N_{\boldsymbol{s}}$ denotes the normalization constant.
	
	\subsection{Eigenstate correction}\label{sec3-3}
	
	On top of the unperturbed part in Eq.~\eqref{eq:U0_solutions}, the leading order correction to eigenstates for the model in Eqs.~\eqref{eq:UF_general} and \eqref{eq:UF_nop1} can be obtained from Eqs.~\eqref{eq:deltaP} and \eqref{eq:eigcorrection},
	\begin{align}\nonumber
		|\omega_{\boldsymbol{s}, \ell}\rangle 
		&= 
		\frac{1}{\sqrt{N_{\boldsymbol{s}}}} \left[
		|\boldsymbol{s}, \ell\rangle +
		\frac{1}{2\pi i} 
		\oint_{\mathscr{C}(\boldsymbol{s},\ell)} \mathrm{d}z
		\sum_{\boldsymbol{s}', \ell'} \frac{e^{iE_{\boldsymbol{s}',\ell'}}}{z - e^{iE_{\boldsymbol{s}',\ell'}}} 
		|\boldsymbol{s}', \ell'\rangle 
		\right. 
		\\ \label{eq:omega_correction1}
		&
		\left.
		\quad 
		\cdot 
		\langle \boldsymbol{s}', \ell'|  \left(i\lambda\sum_{j=1}^L \sigma^x_j\right) |\boldsymbol{s}, \ell\rangle 
		\frac{1}{z- e^{iE_{\boldsymbol{s}, \ell}}}\right].
	\end{align}
	The matrix elements are non-vanishing for $\boldsymbol{s}' = \boldsymbol{s}^{[j]}$ that only differ from $\boldsymbol{s}$ by a spin flip at site $j$, namely,
	\begin{align}
		\boldsymbol{s}^{[j]}: \quad 
		s^{[j]}_{k\neq j} = s_k,\quad 
		s^{[j]}_j = -s_j.
	\end{align}
	Their values can be straightforwardly computed using the expressions in Eq.~\eqref{eq:U0_solutions},
	\begin{align}\nonumber
		\langle \boldsymbol{s}', \ell'| \sigma^x_j |\boldsymbol{s}, \ell\rangle
		&= \delta_{\boldsymbol{s}', \boldsymbol{s}^{[j]}}
		\frac{1}{2}\left[
		e^{-ih_js_j} + (-1)^{\ell+\ell'} e^{ih_js_j}
		\right]
		\\ \label{eq:sigmaxelement}
		&=
		\delta_{\boldsymbol{s}', \boldsymbol{s}^{[j]}}
		\times 
		\begin{cases}
			\cos (h_js_j), & \ell = \ell',
			\\
			-i\sin (h_js_j), & \ell = 1-\ell'.
		\end{cases}
	\end{align}
	Further, using residual theorem, the contour integral gives
	\begin{align}\nonumber
		&\frac{1}{2\pi i} \oint_{\mathscr{C} (\boldsymbol{s}, \ell)} \mathrm{d}z \frac{e^{iE_{\boldsymbol{s}^{[j]}, \ell'}}}{z-e^{iE_{\boldsymbol{s}^{[j]}, \ell'}}} \frac{1}{z-e^{iE_{\boldsymbol{s},\ell}}} 
		=
		\left(e^{i(E_{\boldsymbol{s}, \ell}  -E_{\boldsymbol{s}^{[j]}, \ell'}) } - 1\right)^{-1}
		\\
		&= 
		\begin{cases}
			e^{-i\Delta_{\boldsymbol{s},j}} (2i\sin\Delta_{\boldsymbol{s},j})^{-1}, 
			& \ell = \ell',
			\\
			-e^{-i\Delta_{\boldsymbol{s},j}} (2\cos\Delta_{\boldsymbol{s},j})^{-1}, 
			& \ell = 1- \ell',
		\end{cases}
	\end{align}
	where 
	\begin{align}
		E_{\boldsymbol{s},\ell} - E_{\boldsymbol{s}^{[j]},\ell'} = 2\Delta_{\boldsymbol{s},j} + (\ell-\ell')\pi,
	\end{align}
	and $ \Delta_{\boldsymbol{s},j} $ is given in Eq.~\eqref{eq:Delta}.
	Together, the first-order correction to eigenstates reads
	\begin{align}\nonumber
		|\omega_{\boldsymbol{s}, \ell}\rangle &=
		\frac{1}{\sqrt{N_{\boldsymbol{s}}}}
		\left[ |\boldsymbol{s},\ell\rangle + \frac{\lambda}{2} \sum_{j=1}^L 
		\left(
		|\boldsymbol{s}^{[j]}, \ell\rangle 
		\frac{\cos(h_js_j)}{\sin\Delta_{\boldsymbol{s}, j}}
		\right. \right. 
		\\ \label{eq:eig_nondeg}
		&\qquad
		\left. \left.
		-|\boldsymbol{s}^{[j]}, 1-\ell\rangle \frac{\sin(h_js_j)}{\cos\Delta_{\boldsymbol{s},j}}
		\right)
		e^{-i\Delta_{\boldsymbol{s},j}} \right],
	\end{align}
	where the normalization constant is
	\begin{align}\nonumber
		&N_{\boldsymbol{s}} = 1+ \lambda^2L\bar{V}_1^2(\boldsymbol{s}), 
		\\
		&\bar{V}_1^2(\boldsymbol{s})  = \frac{1}{4L} \sum_{j=1}^L \left[
		\frac{\cos^2(h_j)}{ \sin^2(\Delta_{\boldsymbol{s},j})}
		+
		\frac{\sin^2(h_j)}{ \cos^2(\Delta_{\boldsymbol{s},j})}
		\right].
	\end{align}
	
	While rigorous degeneracy does not occur in disordered systems, near-resonant hybridization among eigenstates can exist. From the expression in Eq.~\eqref{eq:eig_nondeg}, we can identify the resonance condition as $\lambda \gtrsim \min\left( \sin(\Delta_{\boldsymbol{s},j}), \cos(\Delta_{\boldsymbol{s},j}) \right)$,
	because in such cases the first-order correction is no longer much smaller than the leading order term. Then, the resonance condition can be written as
	\begin{align}
		\begin{cases}
			|\boldsymbol{s}, \ell\rangle \text{ and } |\boldsymbol{s}^{[j]}, \ell\rangle, 
			&
			\Delta_{\boldsymbol{s},j} = n\pi,
			\\ \label{eq:deg_cond}
			|\boldsymbol{s}, \ell\rangle \text{ and } |\boldsymbol{s}^{[j]}, 1- \ell\rangle, 
			&
			\Delta_{\boldsymbol{s},j} = \left(n+\frac{1}{2}\right)\pi.
		\end{cases}
	\end{align}
	One can then first perform a degenerate-level diagonalization within the subspace $(|\boldsymbol{s},\ell\rangle, |\boldsymbol{s}^{[j]},\ell'\rangle)$ for the Floquet operator $U_F = U_0e^{i\lambda V} \approx U_0 + i\lambda U_0 V + O(\lambda^2)$, and choose the solution with, i.e., $|\boldsymbol{s},\ell\rangle$ being the dominant component,
	\begin{align}
		|\boldsymbol{s},j\rangle = 
		u_{\boldsymbol{s},j}^+ |\boldsymbol{s},\ell\rangle +  u_{\boldsymbol{s},j}^- |\boldsymbol{s}^{[j]}, \ell'\rangle.
	\end{align}
	Specifically, using 
	$U_0 |\boldsymbol{s},\ell\rangle = e^{iE_{\boldsymbol{s},\ell}}  |\boldsymbol{s},\ell\rangle$, $U_0|\boldsymbol{s}^{[j]}, \ell'\rangle = (-1)^{\ell' - \ell} e^{i(E_{\boldsymbol{s},\ell} - 2\Delta_{\boldsymbol{s},j})} |\boldsymbol{s}^{[j]}, \ell'\rangle$, and Eq. ~\eqref{eq:sigmaxelement} for the matrix elements of $V = \sum_{j=1}^L \sigma^x_j$, we can expand $U_F$ as
	\begin{align}\nonumber
		&\ell = \ell ': \quad \text{subspace } (|\boldsymbol{s},\ell\rangle, |\boldsymbol{s}^{[j]}, \ell\rangle)
		\\ \nonumber
		&e^{i(E_{\boldsymbol{s},\ell}-\Delta_{\boldsymbol{s},j})}
		\begin{pmatrix}
			e^{i\Delta_{\boldsymbol{s},j}} & i\lambda e^{i\Delta_{\boldsymbol{s},j}}\cos(h_js_j) \\
			i\lambda e^{-i\Delta_{\boldsymbol{s},j}} \cos(h_js_j) & e^{-i\Delta_{\boldsymbol{s},j}}
		\end{pmatrix},
		\\ \nonumber
		&\ell = 1-\ell': \quad \text{subspace } (|\boldsymbol{s},\ell\rangle, |\boldsymbol{s}^{[j]}, 1-\ell\rangle)
		\\ \label{eq:deg_mat}
		& 
		e^{i(E_{\boldsymbol{s},\ell}-\Delta_{\boldsymbol{s},j})}
		\begin{pmatrix}
			e^{i\Delta_{\boldsymbol{s},j}} & i\lambda e^{i\Delta_{\boldsymbol{s},j}}i\sin(h_js_j) \\
			i\lambda (-e^{-i\Delta_{\boldsymbol{s},j}}) (-i)\sin(h_js_j) & -e^{-i\Delta_{\boldsymbol{s},j}}
		\end{pmatrix}.
	\end{align}
	Thus, the solution for $\ell=\ell'$ is
	\begin{align}\nonumber
		u_{\boldsymbol{s},j}^{\pm} &= e^{\pm i\Delta_{\boldsymbol{s},j}/2}\sqrt{\frac{1}{2} \left(
			1\pm \frac{\sin\Delta_{\boldsymbol{s},j}}{\sqrt{ \sin^2\Delta_{\boldsymbol{s},j}+ \lambda^2\cos^2(h_js_j)}}
			\right)},
		\\ \label{eq:xsj}
		X_{\boldsymbol{s},j} &\equiv 
		\frac{u^-_{\boldsymbol{s},j}}{u^+_{\boldsymbol{s},j}} = 
		e^{-i\Delta_{\boldsymbol{s},j}} 
		\frac{\sin\Delta_{\boldsymbol{s},j}}{\lambda \cos (h_js_j)}
		\left(
		\sqrt{1+ \left(\frac{\lambda \cos (h_js_j)}{\sin\Delta_{\boldsymbol{s},j}}\right)^2} -1
		\right),
	\end{align}
	while for $\ell = 1-\ell'$,
	\begin{align}\nonumber
		u^\pm_{\boldsymbol{s},j} &= \pm e^{\pm i\Delta_{\boldsymbol{s},j}/2} \sqrt{\frac{1}{2} \left(
			1\pm \frac{\cos\Delta_{\boldsymbol{s},j}}{\sqrt{ \cos^2\Delta_{\boldsymbol{s},j}+ \lambda^2\sin^2(h_js_j)}}
			\right)},
		\\ \label{eq:ysj}
		Y_{\boldsymbol{s},j} &\equiv 
		\frac{u^-_{\boldsymbol{s},j}}{u^+_{\boldsymbol{s},j}} = 
		-e^{-i\Delta_{\boldsymbol{s},j}} 
		\frac{\cos\Delta_{\boldsymbol{s},j}}{\lambda \sin (h_js_j)}
		\left(
		\sqrt{1+ \left(\frac{\lambda \sin (h_js_j)}{\cos\Delta_{\boldsymbol{s},j}}\right)^2} -1
		\right).
	\end{align}
	It is worth noting that in the non-resonant limit, the above result reduces to 
	\begin{align}\nonumber
		X_{\boldsymbol{s},j} &
		\xrightarrow{\lambda\ll \sin(\Delta_{\boldsymbol{s},j})}
		e^{-i\Delta_{\boldsymbol{s},j}} 
		\frac{\lambda}{2} \frac{\cos (h_js_j)}{\sin\Delta_{\boldsymbol{s},j}}
		+O(\lambda^3),
		\\ \label{eq:non_reso}
		Y_{\boldsymbol{s},j} &
		\xrightarrow{\lambda \ll \cos(\Delta_{\boldsymbol{s},j})} 
		- e^{-i\Delta_{\boldsymbol{s},j}} \frac{\lambda}{2} \frac{\sin (h_js_j)}{\cos\Delta_{\boldsymbol{s},j}}
		+ O(\lambda^3),
	\end{align}
	which is exactly the coefficients appearing in Eq.~\eqref{eq:eig_nondeg} for the non-degenerate perturbation corrections. Meanwhile, in the fully degenerate limit, $\Delta_{\boldsymbol{s},j} = n\pi/2$ as in Eq.~\eqref{eq:deg_cond}, we have $ |X_{\boldsymbol{s},j}|, |Y_{\boldsymbol{s},j}| \rightarrow 1$, featuring equal amplitudes between  $|\boldsymbol{s},\ell\rangle$ and the resonant eigenstates $|\boldsymbol{s}^{[j]},\ell\rangle$, $|\boldsymbol{s}^{[j]},1-\ell\rangle$. Thus, we can incorporate the degenerate level hybridization with the non-degenerate perturbative results in Eq.~\eqref{eq:eig_nondeg}, and obtain the first-order corrected eigenstates as
	\begin{align}\nonumber
		|\omega_{\boldsymbol{s},\ell}\rangle &= \frac{1}{\sqrt{N_{\boldsymbol{s}}}} \left[ 
		|\boldsymbol{s}, \ell\rangle 
		+ \sum_{j=1}^L \left(
		X_{\boldsymbol{s},j} |\boldsymbol{s}^{[j]}, \ell\rangle 
		+
		Y_{\boldsymbol{s},j} |\boldsymbol{s}^{[j]}, 1-\ell\rangle
		\right)
		\right],
		\\ \label{eq:eig_corrected}
		&\eta_{\boldsymbol{s}} =  \sum_{j=1}^L\left( |X_{\boldsymbol{s},j}|^2 + |Y_{\boldsymbol{s},j}|^2 \right), 
		\qquad
		N_{\boldsymbol{s}} = 1 + \eta_{\boldsymbol{s}},
	\end{align}
	with $X_{\boldsymbol{s},j}, Y_{\boldsymbol{s},j}$ given in Eqs.~\eqref{eq:xsj} and \eqref{eq:ysj}.

	\subsection{Quasienergy corrections}\label{sec3-4}

	We next consider the quasienergy correction, and especially compare the corrections to each pair of eigenstates with the same spin pattern but different spectral-pairing quantum numbers, namely,  $ |\omega_{\boldsymbol{s},1} - \omega_{\boldsymbol{s}, 0}| - \pi $ in Eq.~\eqref{eq:sp}. In the unperturbed limit, $ |E_{\boldsymbol{s},1} - E_{\boldsymbol{s},0}| = \pi$, while the perturbation theory would show that corrections to $E_{\boldsymbol{s},1}$ and $E_{\boldsymbol{s},0}$ at {\em each} order are exactly identical for all lower-orders $\lambda^{n<L/n_{\text{op}}}$, where the perturbation $V$ in $e^{i\lambda V}$ involves up to $n_{\text{op}}$-spin terms. For the following analysis, it is useful to define the Hamming distance,
	\begin{align}
		{\cal D} \left(\boldsymbol{s}, \boldsymbol{s}' \right) = \frac{1}{2} \sum_{j=1}^L \left|s_j - s_j' \right|,
	\end{align}
	which counts the number of different spins in two Fock states.
	
	From Eq.~\eqref{eq:qecorrection}, the corrected quasienergy $\omega_{\boldsymbol{s},\ell}$ is related to the unperturbed one $E_{\boldsymbol{s},\ell}$ as
	\begin{align}
		\nonumber
		&e^{i\omega_{\boldsymbol{s},\ell}} - e^{iE_{\boldsymbol{s},\ell}}
		=\sum_{n=1}^\infty \text{Tr}\left[
		\delta\Omega_n(\boldsymbol{s},\ell)
		\right],
		\\ \label{eq:deltaOmega}
		&
		\delta\Omega_n(\boldsymbol{s},\ell) = 
		\oint_{\mathscr{C}(\boldsymbol{s},\ell)} \frac{dz}{2\pi i}
		\left(z-e^{iE_{\boldsymbol{s},\ell}} \right) 
		G_0(z) \left[
		U_0 (e^{i\lambda V} -1) G_0(z)
		\right]^n.
	\end{align}
	Taking trace in the unperturbed eigenbases  $|\boldsymbol{s}_0,\ell_0\rangle$ given in Eq.~\eqref{eq:U0_solutions}, we have
	\begin{align} \nonumber
		&\text{Tr}\left[\delta\Omega_n(\boldsymbol{s},\ell)\right]
		= 
		\oint_{\mathscr{C}(\boldsymbol{s},\ell)}
		\frac{dz}{2\pi i} 
		\left(
		z-e^{iE_{\boldsymbol{s},\ell}}
		\right)
		\sum_{\boldsymbol{s}_0,\ell_0}
		\frac{e^{iE_{\boldsymbol{s}_0,\ell_0}}}{ \left(z - e^{iE_{\boldsymbol{s}_0,\ell_0}}\right)^2}
		\\ \label{eq:temp3}
		& \cdot 
		\langle \boldsymbol{s}_0,\ell_0| \left[(e^{i\lambda V}-1)G_0(z)U_0 \right]^{n-1} (e^{i\lambda V}-1) |\boldsymbol{s}_0,\ell_0\rangle .
	\end{align}
	Note that $(e^{i\lambda V}-1) = \sum_{n=1}^\infty \frac{i^m}{m!} \lambda^m V^m$, the matrix elements for terms proportional to $\lambda^M$ are of the form
	\begin{align} \nonumber
		&
		\langle\boldsymbol{s}_0, \ell_0| (V^{m_1} G_0U_0) (V^{m_2}G_0U_0)\dots G_0U_0V^{m_k}|\boldsymbol{s}_0, \ell_0\rangle,
		\\ \label{eq:aM}
		& m_1+m_2+\dots+m_k = M.
	\end{align}
	Inserting completeness relation $1=\sum_{\boldsymbol{s},\ell} |\boldsymbol{s},\ell\rangle \langle\boldsymbol{s},\ell|$ into Eq.~\eqref{eq:aM}, the terms proportional to $\lambda^M$ in Eq.~\eqref{eq:temp3} are
	\begin{align}\nonumber
		&I_M^{\{m_\alpha\}_k}(\boldsymbol{s},\ell) 
		=
		\frac{(i\lambda)^M}{\prod_{\alpha=1}^{k} m_\alpha!}
		\oint_{\mathscr{C}(\boldsymbol{s},\ell)} \frac{dz}{2\pi i} 
		\left( z-e^{iE_{\boldsymbol{s},\ell}} \right)
		\sum_{ \{\boldsymbol{s}_\alpha, \ell_\alpha| \alpha=0,\dots,k-1\}} 
		\\ \nonumber
		&
		\langle \boldsymbol{s}_0,\ell_0| V^{m_1}|\boldsymbol{s}_1,\ell_1\rangle
		\langle \boldsymbol{s}_1,\ell_1| V^{m_2}|\boldsymbol{s}_2,\ell_2\rangle
		\dots
		\langle \boldsymbol{s}_{k-1},\ell_{k-1}| V^{m_k}|\boldsymbol{s}_0,\ell_0\rangle
		\\ \label{eq:temp4}
		&
		\cdot
		\frac{e^{iE_{\boldsymbol{s}_0, \ell_0}}}{\left( z-e^{iE_{\boldsymbol{s}_0,\ell_0}}\right)^2}
		\frac{e^{iE_{\boldsymbol{s}_1, \ell_1}}}{\left( z-e^{iE_{\boldsymbol{s}_1,\ell_1}}\right)}
		\dots
		\frac{e^{iE_{\boldsymbol{s}_{k-1}, \ell_{k-1}}}}{\left( z-e^{iE_{\boldsymbol{s}_{k-1},\ell_{k-1}}}\right)}
		.
	\end{align}
	for all possible combinations of $\{m_\alpha\}$ satisfying Eq.~\eqref{eq:aM}.
	
	In principle, the above integration can be evaluated using the Cauchy's integrals
	\begin{align}
		\oint_{\mathscr{C}(\boldsymbol{s},\ell)}
		\frac{\mathrm{d}z}{2\pi i}
		\frac{f(z)}{(z-e^{iE_{\boldsymbol{s},\ell}})^m} = 
		\left. \frac{d^{m-1}f(z)}{dz^{m-1}}\right|_{z=e^{iE_{\boldsymbol{s},\ell}}},
	\end{align}
	where $f(z)$ is singularity-free at the pole $z=e^{iE_{\boldsymbol{s},\ell}}$ surrounded by the contour $\mathscr{C}(\boldsymbol{s},\ell)$. Nevertheless, it is clear that such integrals sensitively depend on the location of poles, and require factoring out $(z-e^{iE_{\boldsymbol{s},\ell}})^m$ from remaining terms in $f(z)$ . In the expression of Eq.~\eqref{eq:temp4}, there can be $k$ different poles at various locations, making the enumeration cumbersome. To simplify the analysis, we perform a Laurent expansion 
	\begin{align}
		\frac{1}{z-e^{iE_{\boldsymbol{s}_{\alpha},\ell _{\alpha}}}}
		&=
		\sum_{p_\alpha=1}^\infty 
		\frac{\left( -1 \right) ^{p_\alpha-1} 
			\left( z-e^{iE_{\boldsymbol{s},\ell}} \right)^{p_\alpha-1}}{\left( e^{iE_{\boldsymbol{s},\ell}}-e^{iE_{\boldsymbol{s}_{\alpha},\ell _{\alpha}}} \right) ^{p_\alpha}},\nonumber
		\\ \label{eq:Laurent}
		\frac{1}{\left( z-e^{iE_{\boldsymbol{s}_{0},\ell _{0}}} \right) ^2}
		&=
		\sum_{p_0=2}^\infty {\left( p_0-1 \right) \frac{\left( -1 \right) ^{p_0-2}\left( z-e^{iE_{\boldsymbol{s},\ell}} \right) ^{p_0-2}}{\left( e^{iE_{\boldsymbol{s},\ell}}-e^{iE_{\boldsymbol{s}_{0}, \ell _{0}}} \right) ^{p_0}}}.
	\end{align}
	for all $E_{\boldsymbol{s}_\alpha, \ell_\alpha} \neq E_{\boldsymbol{s},\ell}$ around the pole $z=E_{\boldsymbol{s},\ell}$.
	Using Eq.~\eqref{eq:Laurent}, the integrand of Eq.~\eqref{eq:temp4} then only depends on $z$ through the factor $(z-e^{iE_{\boldsymbol{s},\ell}})^m$ of certain integer $m$ concerning one possible pole at $z=e^{iE_{\boldsymbol{s},\ell}}$ alone, while all other factors do not depend on $z$. Then, Cauchy's integrals reduce to the residual theorem
	\begin{align}\label{eq:residual}
		\oint_{\mathscr{C}(\boldsymbol{s},\ell)} 
		\frac{dz}{2\pi i} 
		\sum_{m=-\infty}^\infty C_m(z-e^{iE_{\boldsymbol{s},\ell}})^m
		=
		C_{-1}.
	\end{align}
	Namely, we only need to pick out the factors $C_{-1}$ associated with order-$1$ poles $(z-e^{iE_{\boldsymbol{s},\ell}})^{-1}$, which offers tremendous facilitation for our analysis.

	Based on the Laurent expansion for each term in the denominator of Eq.~\eqref{eq:temp4}, together with the residual theorem in Eq.~\eqref{eq:residual}, a brutal-force calculation leads to the result
	\begin{align}
		\nonumber
		I_M^{\{m_\alpha\}_k}\left( \boldsymbol{s},\ell \right) 
		&=
		\lambda^M \frac{i^M}{\prod_{\alpha=1}^{k} m_\alpha!}
		\sum_{
			\substack{
				p_0,p_1,\dots,p_{k-1}=0,1,\dots
				\\
				p_0+\dots+p_{k-1}=k-1, \,p_0\neq 1
			}
		}^\infty
		{Q_{\left\{ p_{\alpha} \right\}}\left( \boldsymbol{s},\ell \right)},\nonumber
		\\ \nonumber
		Q_{\left\{ p_{\alpha} \right\}}\left( \boldsymbol{s},\ell \right) 
		&=
		C_{\left\{ p_{\alpha} \right\}}
		\sum_{\substack{
				\text{if }p_\alpha\neq 0, \forall \boldsymbol{s}_\alpha,\ell_\alpha \neq \boldsymbol{s},\ell
				\\
				\text{if }p_\alpha= 0, \boldsymbol{s}_\alpha,\ell_\alpha = \boldsymbol{s},\ell\\ 
		}}
		\left[
		\frac{
			\langle \boldsymbol{s}_{k-1},\ell_{k-1}|V^{m_k}|\boldsymbol{s}_0,\ell_0\rangle
		}
		{\left( e^{iE_{\boldsymbol{s},\ell}}-e^{iE_{\boldsymbol{s}_0,\ell _0}} \right)^{p_0}
		}
		e^{iE_{\boldsymbol{s}_{0},\ell _{0}}}
		\right.
		\\ \label{eq:IM_int}
		&\cdot
		\left.
		\prod_{\alpha =1}^{k-1}
		\frac{
			\langle \boldsymbol{s}_{\alpha -1},\ell _{\alpha -1}|V^{m_{\alpha}}|\boldsymbol{s}_{\alpha},\ell _{\alpha}\rangle
		}{
			\left( e^{iE_{\boldsymbol{s},\ell}}-e^{iE_{\boldsymbol{s}_{\alpha},\ell _{\alpha}}} \right) ^{p_{\alpha}}
		}
		e^{iE_{\boldsymbol{s}_{\alpha},\ell _{\alpha}}}
		\right]
		,
	\end{align}
	where the constants
	\begin{align}
		C_{\left\{ p_{\alpha} \right\}}
		&=
		\left( p_0-1 \right) 
		(-1)^{\sum_{\alpha=0}^{k-1} \delta_{p_\alpha,0}}.
	\end{align}
	The constraint for summations over $\{p_\alpha\}$ stems from extracting $\left( z-e^{iE_{\boldsymbol{s},\ell}} \right)^{-1} $ for the pole at $e^{iE_{\boldsymbol{s},\ell}}$ in Eq.~\eqref{eq:temp4},
	\begin{equation}\label{eq:constraint}
		1 + (p_0-2) +\sum_{\alpha =1}^{k-1}{\left( p_{\alpha}-1 \right)}=-1
		\quad\Rightarrow\quad \sum_{\alpha =0}^{k-1}{p_{\alpha}}=k-1.
	\end{equation}
	Also, we have incorporated $p_0,p_1,\dots,p_{k-1} =0$ (i.e., Eq.~\eqref{eq:Laurent} does not involve $p_\alpha=0$) to
	handle the effect of extracting $(z-e^{iE_{\boldsymbol{s},\ell}})^{-1}$ and
	arrive at the compact form of Eq.~\eqref{eq:IM_int}. That affects three aspects for the expression. First, when a certain $p_\alpha=0$, the corresponding factor in the denominator of $Q_{\{p_\alpha\}}(\boldsymbol{s},\ell)$ cancels out $(e^{iE_{\boldsymbol{s},\ell}} - e^{iE_{\boldsymbol{s}_\alpha,\ell_\alpha}})^{p_\alpha=0}|_{\boldsymbol{s}_\alpha, \ell_\alpha = \boldsymbol{s},\ell} = 1$. Second, the summation of $\boldsymbol{s}_\alpha, \ell_\alpha$ is carried out over all possible choices other than $\boldsymbol{s},\ell$ if $p_\alpha\neq0$, while $\boldsymbol{s}_\alpha, \ell_\alpha$ is fixed to $\boldsymbol{s},\ell$ without performing summation if $p_\alpha=0$. Finally, for those $\boldsymbol{s}_\alpha, \ell_\alpha = \boldsymbol{s},\ell$, the corresponding factors $(z-e^{iE_{\boldsymbol{s}_\alpha,\ell_\alpha}})^{-1}$ in Eq.~\eqref{eq:temp4} would not be expanded into Laurent series. Compared with the expressions in Eq.~\eqref{eq:Laurent}, we need to append $(-1)^{\delta_{p_\alpha,0}}$ in $C_{\{p_\alpha\}}$ to cancel out the extra factor $(-1)^{p_\alpha-1}|_{p_\alpha=0}=-1$ for $\alpha=1,2,\dots,k$.

	The crucial step now is to relate 
	$Q_{\left\{ p_{\alpha} \right\}}\left( \boldsymbol{s},\ell \right) $ 
	with
	$Q_{\left\{ p_{\alpha} \right\}}\left( \boldsymbol{s},1-\ell \right)$ for all perturbations of lower orders $M < L/n_{\text{op}}$. In doing so, we first note that the quasienergy difference $E_{\boldsymbol{s},\ell} - E_{\boldsymbol{s}',\ell'} = -\sum_{j=1}^L J_j (s_js_{j+1} - s'_j s'_{j+1}) + \pi(\ell-\ell')$ implies that the numerical factors in Eq.~\eqref{eq:IM_int} for $\ell$ and $(1-\ell)$ always only differ by a minus sign,
	\begin{align}
		\nonumber
		&\frac{
			\prod_{\alpha =0}^{k-1}{e^{iE_{\boldsymbol{s}_{\alpha},1-\ell _{\alpha}}}}
		}{
			\prod_{\alpha =0}^{k-1}{\left( e^{iE_{\boldsymbol{s},1-\ell}}-e^{iE_{\boldsymbol{s}_{\alpha},1-\ell _{\alpha}}} \right)^{p_{\alpha}}}
		}=
		\frac{
			\left( -1 \right)^k 
			\prod_{\alpha =0}^{k-1} e^{iE_{\boldsymbol{s}_{\alpha},\ell _{\alpha}}}
		}{
			(-1)^{\sum_{\alpha =0}^{k-1} p_{\alpha}} 
			\prod_{\alpha =0}^{k-1}
			\left( e^{iE_{\boldsymbol{s},\ell}}-e^{iE_{\boldsymbol{s}_{\alpha},\ell _{\alpha}}} 
			\right)^{p_{\alpha}}
		}\nonumber
		\\ \label{eq:Q_minus}
		&=
		-\frac{
			\prod_{\alpha =0}^{k-1}e^{iE_{\boldsymbol{s}_{\alpha},\ell _{\alpha}}}
		}{
			\prod_{\alpha =0}^{k-1}
			\left( e^{iE_{\boldsymbol{s},\ell}}-e^{iE_{\boldsymbol{s}_{\alpha},\ell _{\alpha}}} 
			\right)^{p_{\alpha}}
		}
	\end{align}
	where we used Eq.~\eqref{eq:constraint} in the final step. Meanwhile, we can expand the matrix elements in the numerators, i.e.,
	\begin{align}\nonumber
		&\langle \boldsymbol{s}_{\alpha-1},\ell_{\alpha-1} |V^{m_\alpha}| 
		\boldsymbol{s}_\alpha,\ell_\alpha\rangle 
		\\ \nonumber
		&=
		\frac{1}{2}\left[
		\langle \boldsymbol{s}_{\alpha-1}| V^{m_\alpha}|\boldsymbol{s}_\alpha\rangle
		+
		(-1)^{\ell_{\alpha}-\ell_{\alpha-1}}
		\langle -\boldsymbol{s}_{\alpha-1}| V^{m_\alpha}|-\boldsymbol{s}_\alpha\rangle 
		\right]
		\\ \label{eq:temp6}
		& +
		\frac{1}{2}\left[
		(-1)^{\ell_{\alpha-1}} 
		\langle -\boldsymbol{s}_{\alpha-1}| V^{m_\alpha}|\boldsymbol{s}_\alpha\rangle 
		+
		(-1)^{\ell_{\alpha}} 
		\langle \boldsymbol{s}_{\alpha-1}| V^{m_\alpha}|-\boldsymbol{s}_\alpha\rangle 
		\right],
	\end{align}
	where we have absorbed the longitudinal fields $e^{i\sum_{j=1}^L h_js_j}$ into the definition of Fock state for conciseness, as they do not affect the proof. For a local perturbation $V$ with up to $n_{\text{op}}$-spin terms, $V^{m_1}$ can only flip up to $n_{\text{op}} m_1$ spins, namely, the Hamming distance between Fock states are bounded,
	\begin{align}\label{eq:necessary}
		\langle \boldsymbol{s}_{\alpha-1}|V^{m_\alpha}|\boldsymbol{s}_\alpha\rangle \neq0 
		\quad \Rightarrow
		{\cal D}(\boldsymbol{s}_{\alpha-1}, \boldsymbol{s}_\alpha) \leqslant n_{\text{op}}m_\alpha.
	\end{align}
	Similarly, we can see that the necessary condition for the first and second line of Eq.~\eqref{eq:temp6} to be non-vanishing are
	\begin{align}\label{eq:d1}
		{\cal D}(\boldsymbol{s}_{\alpha-1}, \boldsymbol{s}_{\alpha})
		&= 
		{\cal D}(-\boldsymbol{s}_{\alpha-1}, 
		-\boldsymbol{s}_{\alpha} ) \leqslant n_{\text{op}}m_{\alpha},
		\\ \label{eq:d2}
		{\cal D}(\boldsymbol{s}_{\alpha-1}, 
		-\boldsymbol{s}_{\alpha})
		&= 
		{\cal D}( -\boldsymbol{s}_{\alpha-1} , \boldsymbol{s}_{\alpha}) \leqslant n_{\text{op}}m_{\alpha},
	\end{align}
	respectively. We shall prove that Eq.~\eqref{eq:d1} and Eq.~\eqref{eq:d2} cannot simultaneously be satisfied for $M<L/n_{\text{op}}$, and therefore either the first or the second line of Eq.~\eqref{eq:temp6} vanishes. In both cases, matrix element only depends on the difference $(-1)^{\ell_{\alpha}-\ell_{\alpha-1}}$. 
	
	Without losing generality, we choose the convention ${\cal D}(\boldsymbol{s}_\alpha, \boldsymbol{s}_{\alpha+1}) \leqslant {\cal D}(\boldsymbol{s}_\alpha, -\boldsymbol{s}_{\alpha+1} )$ to label the cat eigenstates. Then, the numerator in Eq.~\eqref{eq:IM_int} is only non-vanishing if the set of $\{\boldsymbol{s}_\alpha, \ell_\alpha\}$ all satisfy the condition as in Eq.~\eqref{eq:d1},
	\begin{equation}\label{eq:matrix_element_exist}
		\mathcal{D} \left( \boldsymbol{s}_{\alpha} ,\boldsymbol{s}_{\alpha +1} \right) \leqslant m_{\alpha +1}n_\mathrm{op},\quad  \alpha = 0,1,...,k-1,
	\end{equation}
	where we denote $|\boldsymbol{s}_k\rangle \equiv |\boldsymbol{s}_0\rangle$. Then, if in any matrix element, denoted by the subscript $q$, the Fock states simultaneously satisfy the type of condition in Eq.~\eqref{eq:d2}, 
	\begin{align}\label{eq:q_violate}
		{\cal D}(\boldsymbol{s}_q, -\boldsymbol{s}_{q+1}  ) \leqslant m_{q+1} n_{\text{op}},
	\end{align}
	we could repeatedly apply the triangular inequality for Hamming distance
	\begin{align}
		{\cal D}(\boldsymbol{s}_\alpha , \boldsymbol{s}_{\alpha+1} )
		+
		{\cal D}(\boldsymbol{s}_{\alpha+1} , \boldsymbol{s}_{\alpha+2} )
		\geqslant
		{\cal D}(\boldsymbol{s}_\alpha , \boldsymbol{s}_{\alpha+2} )
	\end{align}
	and obtain
	\begin{align}\nonumber
		Mn_{\text{op}} 
		&= 
		\sum_{\alpha=1}^k m_\alpha n_{\text{op}} 
		\geqslant
		{\cal D}(\boldsymbol{s}_q , -\boldsymbol{s}_{q+1} )
		+
		\sum_{\alpha\neq q} 
		{\cal D}(\boldsymbol{s}_\alpha , \boldsymbol{s}_{\alpha+1} )
		\\ \nonumber
		&=
		{\cal D}(\boldsymbol{s}_{q+1} , \boldsymbol{s}_{q+2} ) 
		+
		{\cal D}(\boldsymbol{s}_{q+2} , \boldsymbol{s}_{q+3} ) 
		+
		\dots 
		+
		{\cal D}(\boldsymbol{s}_{k-1} , \boldsymbol{s}_0 )  
		\\ \nonumber
		&\quad +
		{\cal D}(\boldsymbol{s}_0 , \boldsymbol{s}_1 )  
		+\dots +
		{\cal D}(\boldsymbol{s}_{q-1} , \boldsymbol{s}_q )  
		+
		{\cal D}(\boldsymbol{s}_{q} , -\boldsymbol{s}_{q+1} )  
		\\ \label{eq:inequality}
		&\geqslant
		{\cal D}(\boldsymbol{s}_{q+1} , -\boldsymbol{s}_{q+1} )  =L.
	\end{align}
	Thus, for $M<L/n_{\text{op}}$, Eq.~\eqref{eq:inequality} cannot be satisfied, which invalidates Eq.~\eqref{eq:q_violate}. That is, Eq.~\eqref{eq:matrix_element_exist} rules out the possibility for any matrix element to have cross terms obeying Eq.~\eqref{eq:q_violate} for all lower-order perturbations $\lambda^{M<L/n_{\text{op}}}$, i.e., the second line in Eq.~\eqref{eq:temp6} vanishes. 
	One can visualize this relation by thinking of the matrix elements in Eq.~\eqref{eq:IM_int} as a scattering process among Fock states
	\begin{align}
		\boldsymbol{s}_0 \xrightarrow{V^{m_1}}
		\boldsymbol{s}_1 \xrightarrow{V^{m_2}}
		\dots 
		\xrightarrow{V^{m_{k-1}}} \boldsymbol{s}_{k-1}
		\xrightarrow{V^{m_{k}}} \boldsymbol{s}_0,
	\end{align}
	which forms a loop in the Fock space. If another relation simultaneously holds with one or more $\boldsymbol{s}_q$ replaced by $-\boldsymbol{s}_q$, it necessarily means that $V^{m_1+\dots+m_{k}}$ can connect opposite Fock states, namely, $n_{\text{op}}M\geqslant L $. This is only possible for perturbation of orders $O(\lambda^{L/n_{\text{op}}})$. Thus, we have proved that the matrix elements for 
	$Q_{\left\{ p_{\alpha}  \right\}}\left( \boldsymbol{s},\ell \right) $ 
	in Eq.~\eqref{eq:IM_int} only depend on the difference $(-1)^{\ell_{\alpha}-\ell_{\alpha-1}}$ for all lower-order terms $ \lambda^{M<L/n_{\text{op}}}$. In view of $(-1)^{\ell_{\alpha}-\ell_{\alpha-1}} = (-1)^{(1-\ell_{\alpha})-(1-\ell_{\alpha-1})}$, we have for $M = \sum_{\alpha=1}^k m_\alpha <L/n_{\text{op}}$ the relations
	\begin{align}
		\label{eq:l01}
		&
		\langle\boldsymbol{s}_\alpha,\ell_\alpha |V^{m_{\alpha+1}} |\boldsymbol{s}_{\alpha+1},\ell_{\alpha+1}\rangle 
		=
		\langle\boldsymbol{s}_\alpha,1-\ell_\alpha |V^{m_{\alpha+1}} |\boldsymbol{s}_{\alpha+1},1-\ell_{\alpha+1}\rangle.
	\end{align}
	
	In sum, putting the constraints in Eqs.~\eqref{eq:Q_minus} and \eqref{eq:l01} back to Eq.~\eqref{eq:IM_int}, $Q_{\left\{ p_{\alpha} \right\}}\left( \boldsymbol{s},\ell \right) $ of different spectral pairing quantum numbers can be related as
	\begin{widetext}

		\begin{align}\nonumber
			Q_{\left\{ p_{\alpha} \right\}}\left( \boldsymbol{s},1-\ell \right) &
			=C_{\left\{ p_{\alpha} \right\}}
			\sum_{\substack{
					\text{if }p_\alpha\neq 0, \forall \boldsymbol{s}_\alpha,\ell_\alpha \neq \boldsymbol{s},1-\ell
					\\
					\text{if }p_\alpha= 0, \boldsymbol{s}_\alpha,\ell_\alpha = \boldsymbol{s},1-\ell\\ 
			}}
			\frac{\prod_{\alpha =0}^{k-1}{e^{iE_{\boldsymbol{s}_{\alpha},\ell _{\alpha}}}}}{\prod_{\alpha =0}^k{\left( e^{iE_{\boldsymbol{s},\ell}}-e^{iE_{\boldsymbol{s}_{\alpha},\ell _{\alpha}}} \right) ^{p_{\alpha}}}}\prod_{\alpha =1}^k{\langle \boldsymbol{s}_{\alpha -1},\ell _{\alpha -1}|V^{m_{\alpha}}|\boldsymbol{s}_{\alpha},\ell _{\alpha}\rangle}
			\\
			\nonumber
			&=
			C_{\left\{ p_{\alpha} \right\}}
			\sum_{\substack{
					\text{if }p_\alpha\neq 0, \forall \boldsymbol{s}_\alpha,\ell_\alpha \neq \boldsymbol{s},\ell
					\\
					\text{if }p_\alpha= 0, \boldsymbol{s}_\alpha,\ell_\alpha = \boldsymbol{s},\ell\\ 
			}}
			\frac{\prod_{\alpha =0}^{k-1}{e^{iE_{\boldsymbol{s}_{\alpha},1-\ell _{\alpha}}}}}{\prod_{\alpha =0}^k{\left( e^{iE_{\boldsymbol{s},1-\ell}}-e^{iE_{\boldsymbol{s}_{\alpha},1-\ell _{\alpha}}} \right) ^{p_{\alpha}}}}
			\prod_{\alpha =1}^k{\langle \boldsymbol{s}_{\alpha -1},1-\ell _{\alpha -1}|V^{m_{\alpha}}|\boldsymbol{s}_{\alpha},1-\ell _{\alpha}\rangle}
			\\
			&=
			C_{\left\{ p_{\alpha} \right\}}
			\sum_{\substack{
					\text{if }p_\alpha\neq 0, \forall \boldsymbol{s}_\alpha,\ell_\alpha \neq \boldsymbol{s},1-\ell
					\\
					\text{if }p_\alpha= 0, \boldsymbol{s}_\alpha,\ell_\alpha = \boldsymbol{s},1-\ell\\ 
			}}
			\left[ -
			\frac{
				\prod_{\alpha =0}^{k-1}{e^{iE_{\boldsymbol{s}_{\alpha},\ell _{\alpha}}}}
			}{
				\prod_{\alpha =0}^k\left( e^{iE_{\boldsymbol{s},\ell}}-e^{iE_{\boldsymbol{s}_{\alpha},\ell _{\alpha}}} \right)^{p_{\alpha}}
			}
			\right]
			\prod_{\alpha =1}^k{\langle \boldsymbol{s}_{\alpha -1},\ell _{\alpha -1}|V^{m_{\alpha}}|\boldsymbol{s}_{\alpha},\ell _{\alpha}\rangle}
			\nonumber
			\\ \label{eq:Q_relation}
			&=-Q_{\left\{ p_{\alpha} \right\}}\left( \boldsymbol{s},\ell \right) ,
			\qquad
			M<\frac{L}{n_{\text{op}}},
		\end{align}
		
	\end{widetext}
	where we have changed the dummy indices $\ell_\alpha \rightarrow 1-\ell_\alpha$ in the second step, and applied Eq.~\eqref{eq:Q_minus} and Eq.~\eqref{eq:l01} to the third step. Dummy subscript $0$ is replaced with $k$ to simplify the expression, i.e., $|\boldsymbol{s}_k,\ell _k\rangle \equiv |\boldsymbol{s}_0,\ell _0\rangle $. Eq.~\eqref{eq:Q_relation} immediately means
	\begin{align}\label{eq:IM_sp}
		I_{M}^{\{m_\alpha\}_k}(\boldsymbol{s},\ell) = - I_{M}^{\{m_\alpha\}_k}(\boldsymbol{s}, 1-\ell),\quad M<\frac{L}{n_{\text{op}}},
	\end{align}
	for all possible combinations $\{m_\alpha \in\mathbb{N}_+| m_1+m_2+\dots+m_k = M, k= 1,2,\dots, M\}$ in Eq.~\eqref{eq:IM_int}.
	
	Finally, we can feed the result of Eq.~\eqref{eq:IM_sp} back to the quasienergy corrections in Eq.~\eqref{eq:deltaOmega}, where $I_M^{\{m_\alpha\}_k}(\boldsymbol{s},\ell)$ are $\lambda^M$-th order contributions to the right-hand-side of Eq.~\eqref{eq:deltaOmega}, 
	\begin{align} \nonumber
		e^{i\omega_{\boldsymbol{s},\ell}} 
		&= 
		e^{iE_{\boldsymbol{s},\ell}} + \{I_M^{\{m_\alpha\}_k}(\boldsymbol{s}, \ell)| \forall M\}
		\\ \nonumber
		&=
		-\left(
		e^{iE_{\boldsymbol{s},1-\ell}} + \{I_M^{\{m_\alpha\}_k}(\boldsymbol{s}, 1-\ell)| M<\frac{L}{n_{\text{op}}}\}
		\right) + O(\lambda^{M\geqslant L/n_{\text{op}}})
		\\
		&=
		e^{i \left( \omega_{\boldsymbol{s},1-\ell} + \pi + O(\lambda^{L/n_{\text{op}}}) \right)} .
	\end{align}
	Thus, we have the scaling of spectral pairing for systems of size $L$ and perturbations with operator product order $n_{\text{op}}$ as presented in Eq.~\eqref{eq:sp} of Sec.~\ref{sec2}.

	\subsection{Physical observables}\label{sec3-5}
	
	Equipped with the eigenstructure corrections, we proceed to calculate the essential physical observables in DTCs, namely, the IPR in Eq.~\eqref{eq:ipr}, Edwards-Anderson parameters in Eq.~\eqref{eq:EA_ana}, mutual information in Eqs.~\eqref{eq:MI_def} -- \eqref{eq:S2}, and autocorrelators in Eq.~\eqref{eq:autocorr}.
	
	The IPR of a corrected eigenstate $|\omega_{\boldsymbol{s},\ell}\rangle$ can be straightforwardly obtained from the explicit form in Eq.~\eqref{eq:eig_corrected} and the unperturbed solutions in Eq.~\eqref{eq:U0_solutions}, 
	\begin{align}\nonumber
		&\text{IPR}(\omega_{\boldsymbol{s},\ell}) = \sum_{\boldsymbol{s}'} \left|
		\langle\boldsymbol{s}'| \omega_{\boldsymbol{s},\ell}\rangle
		\right|^4
		= \frac{1}{4N_{\boldsymbol{s}}^2}
		\Big[
		|\langle\boldsymbol{s}|\boldsymbol{s}\rangle|^4 + |\langle-\boldsymbol{s}|-\boldsymbol{s}\rangle|^4 
		\\ \nonumber
		& +
		\sum_{j=1}^L \left(
		\left|\langle \boldsymbol{s}^{[j]}|\boldsymbol{s}^{[j]}\rangle (X_{\boldsymbol{s},j} + Y_{\boldsymbol{s},j}) \right|^4 
		+
		\left| \langle - \boldsymbol{s}^{[j]}|-\boldsymbol{s}^{[j]}\rangle (X_{\boldsymbol{s},j} - Y_{\boldsymbol{s},j}) \right|^4 
		\right)
		\Big]
		\\
		&=
		\frac{1}{2N_{\boldsymbol{s}}^2} \left[ 1+ \sum_{j=1}^L \left( 
		|X_{\boldsymbol{s},j}|^4 
		+ |Y_{\boldsymbol{s},j}|^4 
		+ 6 |X_{\boldsymbol{s},j}|^2 |Y_{\boldsymbol{s},j}|^2
		\right)
		\right],
	\end{align}
	which is the result given in Eq.~\eqref{eq:ipr}.

	Further, the Edwards-Anderson parameter for spin glass order defined in Eq.~\eqref{eq:EA_def} can be similarly calculated using the corrected eigenstates of Eq.~\eqref{eq:eig_corrected}.
	Specifically, using $\sigma^z_j \sigma^z_k |\boldsymbol{s},\ell\rangle = s_js_k|\boldsymbol{s}, \ell\rangle$, we have
	\begin{align} \nonumber
		& 
		\langle\omega_{\boldsymbol{s}, \ell}| \sigma^z_j \sigma^z_k | \omega_{\boldsymbol{s}, \ell} \rangle 
		\\ \nonumber
		&= 
		\frac{s_js_k}{N_{\boldsymbol{s}}} 
		\left(
		1 + \eta_{\boldsymbol{s}} - 2(|X_{\boldsymbol{s},j}|^2 + |Y_{\boldsymbol{s},j}|^2 
		+ |X_{\boldsymbol{s},k}|^2 + |Y_{\boldsymbol{s},k}|^2)  
		\right),
	\end{align}
	where we recall the notation $\eta_{\boldsymbol{s}} = \sum_{j=1}^L (|X_{\boldsymbol{s},j}|^2 + |Y_{\boldsymbol{s},j}|^2 )$, and $N_{\boldsymbol{s}} = 1+\eta_{\boldsymbol{s}}$ in Eq.~\eqref{eq:eig_corrected}.
	Then, each term in the summation becomes
	\begin{align}\nonumber
		\langle\omega_{\boldsymbol{s}, \ell}| \sigma^z_j \sigma^z_k | \omega_{\boldsymbol{s}, \ell} \rangle^2
		&=
		1 - \frac{4}{N_{\boldsymbol{s}}} (|X_{\boldsymbol{s},j}|^2 + |Y_{\boldsymbol{s},j}|^2 
		+ |X_{\boldsymbol{s},k}|^2 + |Y_{\boldsymbol{s},k}|^2)
		\\
		&\quad + 
		\frac{4}{N_{\boldsymbol{s}}^2} \left(
		|X_{\boldsymbol{s},j}|^2 + |Y_{\boldsymbol{s},j}|^2 
		+ |X_{\boldsymbol{s},k}|^2 + |Y_{\boldsymbol{s},k}|^2
		\right)^2.
	\end{align}
	Thus, noting that the summation $\sum_{j,k;j\neq k}1 = L(L-1)$ and $\sum_{j, j\neq k} = (L-1)$, we have the Edwards-Anderson parameter
	\begin{align}\nonumber
		&\chi_{\text{EA}} =  \frac{1}{L-1} \sum_{j,k; j\neq k}  \langle \omega_{\boldsymbol{s},\ell}| \sigma^z_j \sigma^z_k |\omega_{\boldsymbol{s},\ell}\rangle^2
		\\ \nonumber
		&= \frac{L(L-1)}{L-1}  
		- \frac{4}{(L-1)N_{\boldsymbol{s}}} 2 \eta_{\boldsymbol{s}} (L-1)
		\\ \nonumber
		&\quad +
		\frac{4}{(L-1)N_{\boldsymbol{s}}^2} 
		\left[
		(L-1) 2\sum_{j=1}^L \left(
		|X_{\boldsymbol{s},j}|^2 + |Y_{\boldsymbol{s},j}|^2
		\right)^2
		\right. 
		\\ \nonumber
		&\left.
		\qquad + 2 (\eta_{\boldsymbol{s}}^2 - \sum_{j=1}^L (
		|X_{\boldsymbol{s},j}|^2 + |Y_{\boldsymbol{s},j}|^2)^2)
		\right]
		\\ \nonumber
		&=
		L - \left[ 
		\frac{8\eta_{\boldsymbol{s}}}{N_{\boldsymbol{s}}} 
		-
		\frac{8\eta_{\boldsymbol{s}}^2}{(L-1)N_{\boldsymbol{s}}^2}
		-
		\frac{8}{N_{\boldsymbol{s}}^2}
		\frac{L-2}{L-1}
		\sum_{j=1}^L \left(
		|X_{\boldsymbol{s},j}|^2 + |Y_{\boldsymbol{s},j}|^2
		\right)^2
		\right] 
		\\ \label{eq:derive_EA}
		&=
		L - \frac{8}{N_{\boldsymbol{s}}} \left[
		\eta_{\boldsymbol{s}} - \frac{\eta_{\boldsymbol{s}}^2 + (L-2) \sum_{j=1}^L \left(
			|X_{\boldsymbol{s},j}|^2 + |Y_{\boldsymbol{s},j}|^2
			\right)^2 }{(L-1)N_{\boldsymbol{s}}} 
		\right],
	\end{align}
	given by Eq.~\eqref{eq:EA_ana}.

	Moving forward to the computation of mutual information in Eq.~\eqref{eq:MI_def}, we need to obtain the reduced density matrix at two far-apart sites $1$ and $L/2$. 
	
	To obtain the reduced density matrix concerning a single site $j=1$ or $L/2$, it is helpful to separate the terms for site $j$ in a corrected eigenstate as
	\begin{align}\nonumber
		|\omega_{\boldsymbol{s},\ell}\rangle &= \frac{1}{\sqrt{N_{\boldsymbol{s}}}} \left(
		|\boldsymbol{s},\ell\rangle 
		+
		X_{\boldsymbol{s},j} |\boldsymbol{s}^{[j]},\ell\rangle  + Y_{\boldsymbol{s},j} |\boldsymbol{s}^{[j]},1-\ell\rangle  
		\right)
		\\ 
		&\quad +
		\frac{1}{\sqrt{N_{\boldsymbol{s}}}} \sum_{m\neq j}
		\left( X_{\boldsymbol{s},m} |\boldsymbol{s}^{[m]},\ell\rangle  + Y_{\boldsymbol{s},m} |\boldsymbol{s}^{[m]},1-\ell\rangle  \right)
	\end{align}
	The single-site reduced density matrix can be expanded into
	\begin{align}\nonumber
		\rho_j &= \text{Tr}_{q\neq j}\left( |\omega_{\boldsymbol{s},\ell} \rangle 
		\langle \omega_{\boldsymbol{s},\ell} | \right)
		= a_j|s_j\rangle \langle s_j| + b_j|s_j\rangle \langle -s_j| 
		\\
		&\qquad 
		+ b^*_j |-s_j\rangle \langle s_j| + (1-a_j)|-s_j\rangle \langle -s_j|.
	\end{align}
	Note the expression of $|\boldsymbol{s},\ell\rangle$ in Eq.~\eqref{eq:U0_solutions}, we have the partial trace over sites $q$ other than $j$ as
	\begin{align}\nonumber
		&
		\text{Tr}_{q\neq j} |\boldsymbol{s},\ell\rangle \langle \boldsymbol{s},\ell'| 
		=
		\frac{1}{2} (|s_j\rangle \langle s_j| + (-1)^{\ell - \ell'} |-s_j\rangle \langle -s_j|)
		\\ \nonumber
		&
		= (-1)^{\ell-\ell'} \text{Tr}_{q\neq j} |\boldsymbol{s}^{[j]},\ell\rangle \langle \boldsymbol{s}^{[j]},\ell'| 
		=
		\text{Tr}_{q\neq j} |\boldsymbol{s}^{[p\neq j]},\ell\rangle \langle \boldsymbol{s}^{[p\neq j]}, \ell' \rangle 
		,
		\\ \nonumber
		&
		\text{Tr}_{q\neq j}|\boldsymbol{s},\ell\rangle \langle \boldsymbol{s}^{[j]},\ell'|
		=
		\frac{1}{2} ( 
		e^{-ih_js_j} |s_j\rangle \langle -s_j| 
		+
		e^{ih_js_j} (-1)^{\ell-\ell'} |-s_j\rangle \langle s_j| ),
		\\ 
		&
		\text{Tr}_{q\neq j} |\boldsymbol{s},\ell\rangle \langle \boldsymbol{s}^{[p\neq j]}, \ell' | 
		= 
		\text{Tr}_{q\neq j} |\boldsymbol{s}^{[j]},\ell\rangle \langle \boldsymbol{s}^{[p\neq j]}, \ell' | = 0.
	\end{align}
	Then, the components in a reduced density matrix can be computed as
	\begin{align}\nonumber
		a_j &= \frac{1}{2N_{\boldsymbol{s}}} \left(
		1+ \sum_{j=1}^L (|X_{\boldsymbol{s},j}|^2 + |Y_{\boldsymbol{s},j}|^2)
		\right)
		\\\nonumber
		& \quad +
		\frac{1}{2N_{\boldsymbol{s}}} \left(
		-X^*_{\boldsymbol{s},j} Y_{\boldsymbol{s},j} - X_{\boldsymbol{s},j} Y_{\boldsymbol{s},j}^* + \sum_{m\neq j} (X^*_{\boldsymbol{s},m} Y_{\boldsymbol{s},m} + X_{\boldsymbol{s},m} Y_{\boldsymbol{s},m}^*) 
		\right)
		\\ \nonumber
		&=
		\frac{1}{2} + \frac{1}{N_{\boldsymbol{s}}}
		\left(
		-\text{Re}(X^*_{\boldsymbol{s},j} Y_{\boldsymbol{s},j}) + \sum_{m\neq j} \text{Re} (X^*_{\boldsymbol{s},m} Y_{\boldsymbol{s},m}) 
		\right),
		\\ \nonumber
		b_j &= \frac{e^{-ih_js_j}}{2N_{\boldsymbol{s}}} \left(  (X_{\boldsymbol{s},j}^* + Y_{\boldsymbol{s},j}^*)
		+ 
		(X_{\boldsymbol{s},j} - Y_{\boldsymbol{s},j})
		\right)
		\\
		&=
		\frac{e^{-ih_js_j}}{N_{\boldsymbol{s}}} \left(
		\text{Re}(X_{\boldsymbol{s},j})
		- i\text{Im}(Y_{\boldsymbol{s},j})
		\right).
	\end{align}
	where we used $\eta_{\boldsymbol{s}} = \sum_{j=1}^L(|X_{\boldsymbol{s},j}|^2 + |Y_{\boldsymbol{s},j}|^2)$ and $N_{\boldsymbol{s}} = 1+\eta_{\boldsymbol{s}}$ from Eq.~\eqref{eq:eig_corrected}. Re$(\dots)$ and Im$(\dots)$ means taking the real and imaginary part, respectively. Then, the single-site entanglement entropy can be obtained from the eigenvalues of $\rho_j$ as
	\begin{align}\nonumber
		S_j &= -p_j^+ \ln p_j^+ - p_j^-\ln p_j^-,
		\\
		p_j^\pm &= \frac{1}{2} \pm \sqrt{(a_j-1/2)^2 + |b_j|^2},
	\end{align}
	which gives the results $S_1, S_{L/2}$ in Eq.~\eqref{eq:S1}. 
	
	For the two-site reduced density matrix $\rho_{j,k}$,
	we can similarly separate out terms concerning the sites $j$ and $k$ in a corrected eigenstate	\begin{align}\nonumber
		|\omega_{\boldsymbol{s},\ell}\rangle &= \frac{|\boldsymbol{s},\ell\rangle }{\sqrt{N_{\boldsymbol{s}}}} 
		+
		\frac{1}{\sqrt{N_{\boldsymbol{s}}}}\sum_{m=j,k} \left(X_{\boldsymbol{s},m} |\boldsymbol{s}^{[m]},\ell\rangle  + Y_{\boldsymbol{s},m} |\boldsymbol{s}^{[m]},1-\ell\rangle  
		\right)
		\\ 
		&\quad +
		\frac{1}{\sqrt{N_{\boldsymbol{s}}}} \sum_{m\neq j,k}
		\left( X_{\boldsymbol{s},m} |\boldsymbol{s}^{[m]},\ell\rangle  + Y_{\boldsymbol{s},m} |\boldsymbol{s}^{[m]},1-\ell\rangle  \right).
	\end{align}
	Using the non-vanishing partial trace results
	\begin{align}\nonumber
		&
		\text{Tr}_{q\neq j,k} |\boldsymbol{s},\ell\rangle \langle \boldsymbol{s}, \ell'| 
		=
		\text{Tr}_{q\neq j,k} |\boldsymbol{s}^{[m\neq j,k]},\ell\rangle \langle \boldsymbol{s}^{[m\neq j,k]}, \ell'| 
		\\ \nonumber
		&= 
		\frac{1}{2}(|s_js_k\rangle\langle s_j s_k| + 
		(-1)^{\ell-\ell'} |-s_j,-s_k\rangle \langle -s_j,-s_k| )
		\\ \nonumber
		&
		\text{Tr}_{q\neq j,k} |\boldsymbol{s}^{[j]},\ell\rangle \langle \boldsymbol{s}^{[j]}, \ell'| 
		=
		(-1)^{\ell-\ell'}\text{Tr}_{q\neq j,k} |\boldsymbol{s}^{[k]},\ell\rangle \langle \boldsymbol{s}^{[k]}, \ell'| 
		\\ \nonumber
		&= 
		\frac{1}{2}(|-s_js_k\rangle\langle -s_j s_k| + 
		(-1)^{\ell-\ell'} |s_j,-s_k\rangle \langle s_j,-s_k| )
		\\ \nonumber
		&\text{Tr}_{q\neq j,k} |\boldsymbol{s},\ell\rangle \langle \boldsymbol{s}^{[j]}, \ell'| 
		\\ \nonumber
		&= 
		\frac{1}{2} (e^{-ih_js_j}|s_js_k\rangle \langle -s_js_k| + (-1)^{\ell-\ell'} e^{ih_js_j} |-s_j,-s_k\rangle \langle s_j,-s_k|)
		\\ \nonumber
		&\text{Tr}_{q\neq j,k} |\boldsymbol{s},\ell\rangle \langle \boldsymbol{s}^{[k]}, \ell'| 
		\\ \nonumber
		&= 
		\frac{1}{2} (e^{-ih_ks_k}|s_j,s_k\rangle \langle s_j,-s_k| + (-1)^{\ell-\ell'} e^{ih_ks_k}|-s_j,-s_k\rangle \langle -s_j,s_k|)
		\\ \nonumber
		&\text{Tr}_{q\neq j,k} |\boldsymbol{s}^{[j]},\ell\rangle \langle \boldsymbol{s}^{[k]}, \ell'| 
		=
		(-1)^{\ell-\ell'}\text{Tr}_{q\neq j,k} |\boldsymbol{s}^{[k]},\ell\rangle \langle \boldsymbol{s}^{[j]}, \ell'| 
		\\ \nonumber
		&= 
		\frac{1}{2} (e^{-i( h_ks_k - h_js_j)}|-s_js_k\rangle \langle s_j,-s_k| \\ 
		&\quad + (-1)^{\ell-\ell'} e^{i( h_ks_k - h_js_j)} |s_j,-s_k\rangle \langle -s_j,s_k|),
	\end{align}
	the coefficients for $\rho_{j,k}$ include the diagonal terms
	\begin{align} \nonumber
		&
		|s_js_k\rangle \langle s_js_k|\times 
		\frac{1}{2N_{\boldsymbol{s}}} \left(
		1 +
		\sum_{m\neq j,k} |X_{\boldsymbol{s},m} + Y_{\boldsymbol{s},m}|^2
		\right),
		\\ \nonumber
		&
		|-s_j,-s_k\rangle \langle -s_j,-s_k|\times 
		\frac{1}{2N_{\boldsymbol{s}}} \left(
		1 +
		\sum_{m\neq j,k} |X_{\boldsymbol{s},m} - Y_{\boldsymbol{s},m}|^2
		\right),
		\\ \nonumber
		&
		|-s_js_k\rangle \langle -s_js_k|\times \frac{1}{2N_{\boldsymbol{s}}} 
		\left(
		|X_{\boldsymbol{s},j} + Y_{\boldsymbol{s},j}|^2 
		+ 
		|X_{\boldsymbol{s},k} - Y_{\boldsymbol{s},k}|^2
		\right),
		\\ 
		&|s_j, -s_k\rangle \langle s_j, -s_k|\times \frac{1}{2N_{\boldsymbol{s}}} 
		\left(
		|X_{\boldsymbol{s},j} - Y_{\boldsymbol{s},j}|^2 
		+ 
		|X_{\boldsymbol{s},k} + Y_{\boldsymbol{s},k}|^2
		\right),
	\end{align}
	and off-diagonal terms
	\begin{align}\nonumber
		&
		|s_js_k\rangle \langle -s_js_k| \times \frac{e^{-ih_js_j}}{2N_{\boldsymbol{s}}} 
		\left(
		X_{\boldsymbol{s},j}^* + Y_{\boldsymbol{s},j}^*
		\right),
		\\ \nonumber
		& |s_js_k\rangle \langle s_j,-s_k| \times  \frac{e^{-ih_ks_k}}{2N_{\boldsymbol{s}}} 
		\left(
		X_{\boldsymbol{s},k}^* + Y_{\boldsymbol{s},k}^*
		\right),
		\\
		\nonumber
		&
		|s_js_k\rangle \langle -s_j,-s_k| \times 0,
		\\
		\nonumber
		&
		|-s_js_k\rangle \langle s_j,-s_k| \times 
		\frac{e^{-i(h_ks_k - h_js_j)}}{2N_{\boldsymbol{s}}} 
		\left[  
		\left(
		X_{\boldsymbol{s},j} + Y_{\boldsymbol{s},j}
		\right)
		\left(
		X_{\boldsymbol{s},k}^* + Y_{\boldsymbol{s},k}^*
		\right)
		\right. 
		\\ \nonumber
		& \left. 
		\qquad
		+ 
		\left(
		X_{\boldsymbol{s},k} - Y_{\boldsymbol{s},k}
		\right)
		\left(
		X_{\boldsymbol{s},j}^* - Y_{\boldsymbol{s},j}^*
		\right)
		\right],
		\\ \nonumber
		&
		|-s_js_k\rangle \langle -s_j,-s_k| \times  \frac{e^{-ih_ks_k}}{2N_{\boldsymbol{s}}} 
		\left(
		X_{\boldsymbol{s},k} - Y_{\boldsymbol{s},k}
		\right),
		\\
		&
		|s_j,-s_k\rangle \langle -s_j,-s_k| \times  \frac{e^{-ih_js_j}}{2N_{\boldsymbol{s}}} 
		\left(
		X_{\boldsymbol{s},j} - Y_{\boldsymbol{s},j}
		\right).
	\end{align}
	Remaining terms can be obtained by using the Hermiticity of the density matrix $\rho_{jk} = \left(\rho_{j,k}\right) ^\dagger$. Then, we have the two-site reduced density matrix and further the entanglement entropy $S_{1,L/2}$ as in Eq.~\eqref{eq:S2}.

	Finally, we consider the dynamics of autocorrelators in Eq.~\eqref{eq:corr}, and relate it to the eigenstructure obtained previously. The disordered DTCs feature local oscillations for arbitrary initial spin patterns, and we would consider the autocorrelators averaged over all possible spin patterns
	\begin{align}
		\langle C(t) \rangle_{\boldsymbol{s}} = \frac{1}{2^L}\sum_{\boldsymbol{s}}  \frac{1}{L} \sum_{j=1}^L 
		\langle \boldsymbol{s}| \sigma^z_j(t) \sigma^z_j(0) |\boldsymbol{s}\rangle, 
	\end{align}
	where $\sigma^z_j(t) = (U_F^\dagger)^{t/T} \sigma^z_j U_F^{t/T}$. In practice, after disorder average, all initial spin patterns are equivalent and would render the same evolution signals. Thus, $ \langle C(t) \rangle_{\boldsymbol{s}}$ is the same as autocorrelators evolved from a randomly sampled initial Fock state $|\boldsymbol{s}\rangle$ averaged over disorder realizations.
	
	To relate $C(t)$ to the eigenstructure, we expand the Floquet operator at $t=nT$ stroboscopic period end in its eigenbases $U_F^{t/T} = e^{i\omega_{\boldsymbol{s},\ell}(t/T)} |\omega_{\boldsymbol{s},\ell}\rangle 
	\langle \omega_{\boldsymbol{s},\ell}| $,
	\begin{align}\nonumber
		&
		\langle C(t)\rangle_{\boldsymbol{s}} = \frac{1}{2^LL} \sum_{\boldsymbol{s}} \sum_{j=1}^L 
		s_j 
		\left[ 
		\sum_{\boldsymbol{s}',\ell'} |\langle \boldsymbol{s}|\omega_{\boldsymbol{s}',\ell'}\rangle |^2 \langle \omega_{\boldsymbol{s}',\ell'}
		|\sigma^z_j| \omega_{\boldsymbol{s}',\ell'}\rangle 
		\right. 
		\\ \label{eq:corr_def}
		& \left. 
		+ \sum_{\boldsymbol{s}', \ell' \neq \boldsymbol{s}'', \ell''} 
		\langle \omega_{\boldsymbol{s}',\ell'} |\sigma^z_j |\omega_{\boldsymbol{s}'',\ell''}\rangle 
		\langle \omega_{\boldsymbol{s}'',\ell''}|\boldsymbol{s}\rangle \langle \boldsymbol{s}| \omega_{\boldsymbol{s}', \ell'}\rangle 
		e^{i(\omega_{\boldsymbol{s}'',\ell''} - \omega_{\boldsymbol{s}', \ell'})t/T}\right],
	\end{align}
	where we separate the terms into diagonal and off-diagonal ones in the expansion. Note that
	$\sum_{\boldsymbol{s}} s_j |\langle \boldsymbol{s}|\omega_{\boldsymbol{s}',\ell'}\rangle|^2 = \langle \omega_{\boldsymbol{s}',\ell'}| \left(
	\sum_{\boldsymbol{s}} s_j|\boldsymbol{s}\rangle \langle \boldsymbol{s}
	\right)| \omega_{\boldsymbol{s}',\ell'}\rangle
	=\langle \omega_{\boldsymbol{s}',\ell'}|\sigma^z_j| \omega_{\boldsymbol{s}',\ell'}\rangle$,
	the diagonal part in the first line of Eq.~\eqref{eq:corr_def} reduces to a time-independent constant $A_0$ for averaged magnetization,
	\begin{align}
		A_0 = \frac{1}{2^LL}\sum_{j=1}^L  \sum_{\boldsymbol{s}', \ell'} \langle \omega_{\boldsymbol{s}', \ell'}| \sigma^z_j|\omega_{\boldsymbol{s}', \ell'}\rangle^2.
	\end{align}
	Meanwhile, the off-diagonal part in the second line of Eq.~\eqref{eq:corr_def} can be similarly reduced to
	\begin{align}
		\frac{1}{2^LL} \sum_{j=1}^L \sum_{\boldsymbol{s}', \ell' \neq \boldsymbol{s}'', \ell''} 
		|\langle \omega_{\boldsymbol{s}',\ell'} |\sigma^z_j |\omega_{\boldsymbol{s}'',\ell''}\rangle |^2 e^{i(\omega_{\boldsymbol{s}'',\ell''} - \omega_{\boldsymbol{s}', \ell'})t/T}.
	\end{align}
	For the majority pairs of $(\boldsymbol{s}',\ell')$ and $(\boldsymbol{s}'', \ell'')$, the perturbed quasienergy difference $(\omega_{\boldsymbol{s}',\ell'} - \omega_{\boldsymbol{s}'',\ell''})$ are random, and therefore one can consider an argument similar to the eigenstate thermalization hypothesis that contributions from these terms at the long time limit vanish due to dephasing. However, for the spectral paired ones $(\omega_{\boldsymbol{s}',\ell'} - \omega_{\boldsymbol{s}',1-\ell'})=\pi + O(\lambda^{L/n_{\text{op}}})$, their quasienergy difference is fixed to $\pi$ with exponential accuracy under generic perturbations, as in Eq.~\eqref{eq:sp}. Thus, we only keep the contributions from spectral-paired ones and obtain the off-diagonal part as
	\begin{align}\nonumber
		& 
		A_1\cos\left[ \left(\pi+O(\lambda^{L/n_{\text{op}}})\right) \frac{t}{T} \right]
		=
		A_1(-1)^{t/T} \cos\left(\frac{t}{T}O(\lambda^{L/n_{\text{op}}})\right),
		\\ \label{eq:temp2}
		&A_1 = 
		\frac{1}{2^LL}\sum_{j=1}^L \sum_{\boldsymbol{s}}
		|\langle \omega_{\boldsymbol{s},0}| \sigma^z_j |\omega_{\boldsymbol{s},1}\rangle|^2 .
	\end{align}
	Thus, the autocorrelator becomes
	\begin{align}\label{eq:cnt}
		\langle C(t) \rangle_{\boldsymbol{s}} = A_0 + A_1 (-1)^{t/T} \cos\left(\frac{t}{T} O(\lambda^{L/n_{\text{op}}})\right).
	\end{align}
	That means the autocorrelator oscillates around $A_0$ with an amplitude of $\pm A_1$ and rigid period $2T$, and such an oscillation is modulated with a slow-varying envelope of exponentially long timescale, which is usually understood as the lifetime of DTC oscillation in a finite-size system,
	\begin{align}
		\tau_* = (1/\lambda)^{L/n_{\text{op}}} T = e^{(L/n_{\text{op}})\times \ln(1/\lambda)} T.
	\end{align}
	
	The static and oscillating amplitudes of autocorrelators $A_0$ and $A_1$ can be evaluated using the corrected eigenstates $|\omega_{\boldsymbol{s},\ell}\rangle$ in Eq.~\eqref{eq:eig_corrected}. Specifically, for the static part $A_0$, note from \eqref{eq:U0_solutions} that the unperturbed ideal cat eigenstates have
	\begin{align}\label{eq:slz}
		&
		\langle \boldsymbol{s},\ell| \sigma^z_j|\boldsymbol{s},\ell'\rangle = \frac{1-(-1)^{\ell-\ell'}}{2} s_j = 
		\begin{cases}
			0, & \ell = \ell',
			\\
			s_j, & \ell = 1-\ell'.
		\end{cases}
	\end{align}
	Thus, we obtain from the corrected eigenstates in Eq.~\eqref{eq:eig_corrected} that
	\begin{align} 
		\langle \omega_{\boldsymbol{s},\ell}| \sigma^z_j |\omega_{\boldsymbol{s}, \ell}\rangle 
		&
		=\frac{2s_j}{N_{\boldsymbol{s}}} \left[ 
		-\text{Re}\left( X_{\boldsymbol{s},j}^* Y_{\boldsymbol{s},j} \right) 
		+
		\sum_{k\neq j}
		\text{Re}\left( X_{\boldsymbol{s},k}^* Y_{\boldsymbol{s},k} \right) 
		\right].
	\end{align}
	Therefore, the static amplitude becomes
	\begin{align}\label{eq:A0}
		A_0 &= \frac{4}{2^L L} \sum_{j=1}^L \sum_{\boldsymbol{s},\ell} 
		\frac{1}{N_{\boldsymbol{s}}^2}
		\left[
		-\text{Re}\left( X_{\boldsymbol{s},j} Y_{\boldsymbol{s},j}^* \right)
		+
		\sum_{k\neq j}
		\text{Re}\left( X_{\boldsymbol{s},k} Y_{\boldsymbol{s},k}^* \right)  \right]^2.
	\end{align}
	It is worth noting that $A_0$ is always of perturbative strength, as $X_{\boldsymbol{s},j}, Y_{\boldsymbol{s},j} \rightarrow0$ when $\lambda\rightarrow0$. This may appear counter-intuitive at first glance, because the unperturbed part $U_0$ can involve strong longitudinal fields $h_j$ in Eq.~\eqref{eq:UF_general}, which may be expected to polarize the spin and result in a large static value for $C(t)$. But different from a static Ising model, DTCs involve an almost perfect $\pi$-pulse, which serves to cancel the effect of longitudinal fields every other periodic, i.e. $U_F^2 \sim (e^{ih_j \sigma^z_j} \sigma^x_j)^2 = e^{ih_j \sigma^z_j} e^{-ih_j \sigma^z_j} = 1 $ as we used $\sigma^x_j \sigma^z_j \sigma^x_j = - \sigma^z_j$. From the perspective of perturbation theory, this is because the unperturbed eigenstates are cat states that are superposition of opposite spin patterns $\left|\pm\boldsymbol{s}\right\rangle$, which rigorously have vanishing net magnetic order $\langle \boldsymbol{s},\ell| \sigma^z_j| \boldsymbol{s}, \ell\rangle = 0$. Thus, any remaining static magnetism results from the perturbed terms in $ |\omega_{\boldsymbol{s},\ell}\rangle $.
	
	On the other hand, the oscillation amplitude can be similarly calculated using Eq.~\eqref{eq:eig_corrected} and~\eqref{eq:slz},
	\begin{align}\nonumber
		&
		\langle \omega_{\boldsymbol{s},0}| \sigma^z_j |\omega_{\boldsymbol{s},1}\rangle 
		\\ \nonumber
		&= 
		\frac{s_j}{N_{\boldsymbol{s}}} 
		\left[ 1
		- \left(|X_{\boldsymbol{s},j}|^2 + |Y_{\boldsymbol{s},j}|^2 \right)
		+
		\sum_{k\neq j}
		\left(|X_{\boldsymbol{s},k}|^2 + |Y_{\boldsymbol{s},k}|^2 \right)
		\right]
		\\
		\nonumber
		&=
		s_j \left[
		1 - \frac{2}{N_{\boldsymbol{s}}}\left(|X_{\boldsymbol{s},j}|^2 + |Y_{\boldsymbol{s},j}|^2 \right)
		\right],
		\\ \nonumber
		&A_1 = \frac{1}{2^LL} \sum_{\boldsymbol{s}} \sum_{j=1}^L  
		\left[
		1 - \frac{2}{N_{\boldsymbol{s}}}\left(|X_{\boldsymbol{s},j}|^2 + |Y_{\boldsymbol{s},j}|^2 \right)
		\right]^2
		\\ \label{eq:A1}
		&
		\quad = 1 -
		\frac{1}{2^L} \sum_{\boldsymbol{s}} 
		\frac{4}{N_{\boldsymbol{s}}L}
		\left[
		\eta_{\boldsymbol{s}}
		-
		\frac{1}{N_{\boldsymbol{s}}} 
		\sum_{j=1}^L
		\left(|X_{\boldsymbol{s},j}|^2 + |Y_{\boldsymbol{s},j}|^2 \right)^2
		\right]
		,
	\end{align}
	where we once again used $\eta_{\boldsymbol{s}} = \sum_{j=1}^L (|X_{\boldsymbol{s},j}|^2 + |Y_{\boldsymbol{s},j}|^2),
	N_{\boldsymbol{s}} = 1+\eta_{\boldsymbol{s}}$.
	Note that in the unperturbed limit $X_{\boldsymbol{s},j}, Y_{\boldsymbol{s},j}\rightarrow 0$, and therefore the oscillation amplitude $A_1 \rightarrow 1$ as expected. Results in Eqs.~\eqref{eq:cnt}, \eqref{eq:A0}, and \eqref{eq:A1} then give the analytical prediction for autocorrelator evolution.
	
	\subsection{Eigenstate correction for two-spin perturbation}\label{sec-2spin}
	
	We now calculate the eigenstate correction for $V = \sum_{j=1}^L (\sigma^x_j + \sigma^x_{j}\sigma^x_{j+1})$, which includes both the single- and two-spin perturbations, following the same procedures as in Sec.~\ref{sec3-3}. Analogous to Eq.~\eqref{eq:omega_correction1}, the general form of corrected eigenstates now reads
	\begin{align}\nonumber
		|\omega_{\boldsymbol{s}, \ell}\rangle 
		&= 
		\frac{1}{\sqrt{N_{\boldsymbol{s}}}} \left[
		|\boldsymbol{s}, \ell\rangle +
		\frac{1}{2\pi i} 
		\oint_{\mathscr{C}(\boldsymbol{s},\ell)} \mathrm{d}z
		\sum_{\boldsymbol{s}', \ell'} \frac{e^{iE_{\boldsymbol{s}',\ell'}}}{z - e^{iE_{\boldsymbol{s}',\ell'}}} 
		|\boldsymbol{s}', \ell'\rangle 
		\right. 
		\\ \label{eq:omega_correction2}
		&
		\left.
		\quad 
		\times 
		\langle \boldsymbol{s}', \ell'|  \left(i\lambda\sum_{j=1}^L (\sigma^x_j + \sigma^x_j \sigma^x_{j+1}) \right) |\boldsymbol{s}, \ell\rangle 
		\frac{1}{z- e^{iE_{\boldsymbol{s}, \ell}}}\right].
	\end{align}
	The minor difference is that the matrix elements relate spin patterns $\boldsymbol{s}$ to both the cases with one or two spins flipped
	\begin{align}\nonumber
		&
		\langle \boldsymbol{s}', \ell'| 
		\sigma^x_j + \sigma^x_j \sigma^x_{j+1}
		|\boldsymbol{s}, \ell\rangle
		= \delta_{\boldsymbol{s}', \boldsymbol{s}^{[j]}}
		\frac{1}{2}\left[
		e^{-ih_js_j} + (-1)^{\ell+\ell'} e^{ih_js_j}
		\right]
		\\ \nonumber
		&+ \delta_{\boldsymbol{s}', \boldsymbol{s}^{[j,j+1]}}
		\frac{1}{2}\left[
		e^{-i(h_js_j + h_{j+1}s_{j+1})} + (-1)^{\ell+\ell'} e^{i(h_js_j + h_{j+1}s_{j+1})}
		\right]
		\\  \nonumber
		&=
		\delta_{\boldsymbol{s}', \boldsymbol{s}^{[j]}}
		\times 
		\begin{cases}
			\cos (h_js_j), & \ell = \ell',
			\\
			-i\sin (h_js_j), & \ell = 1-\ell'.
		\end{cases}
		\\ 
		&\quad +
		\delta_{\boldsymbol{s}', \boldsymbol{s}^{[j,j+1]}}
		\times 
		\begin{cases}
			\cos (h_js_j+h_{j+1}s_{j+1}), & \ell = \ell',
			\\
			-i\sin (h_js_j+h_{j+1}s_{j+1}), & \ell = 1-\ell',
		\end{cases}
	\end{align}
	where the spin pattern differing by two flipped sites reads
	\begin{align} \nonumber
		&
		\boldsymbol{s}^{[j,j+1]}: \quad 
		s_{k\neq j, j+1}^{[j,j+1]} = s_k,
		\quad
		s_{j}^{[j,j+1]} = - s_j, s_{j+1}^{[j,j+1]} = -s_{j+1};
		\\
		&
		\Delta_{\boldsymbol{s},j,j+1} = \frac{ E_{\boldsymbol{s},\ell} - E_{\boldsymbol{s}^{[j,j+1]},\ell} }{2}
		=
		-J_{j-1}s_{j-1}s_j - J_{j+1} s_{j+1}s_{j+2},
	\end{align}
	with $E_{\boldsymbol{s},\ell} = -\sum_{j=1}^L J_j s_j s_{j+1} + \pi\ell$ given by Eq.~\eqref{eq:U0_solutions}.
	Meanwhile, the relevant contour integrals for non-vanishing matrix elements include the following,
	\begin{align}\nonumber
		&\frac{1}{2\pi i} \oint_{\mathscr{C} (\boldsymbol{s}, \ell)} \mathrm{d}z \frac{e^{iE_{\boldsymbol{s}^{[j]}, \ell'}}}{z-e^{iE_{\boldsymbol{s}^{[j]}, \ell'}}} \frac{1}{z-e^{iE_{\boldsymbol{s},\ell}}} 
		=
		\left(e^{i(E_{\boldsymbol{s}, \ell}  -E_{\boldsymbol{s}^{[j]}, \ell'}) } - 1\right)^{-1}
		\\
		&= 
		\begin{cases}
			e^{-i\Delta_{\boldsymbol{s},j}} (2i\sin\Delta_{\boldsymbol{s},j})^{-1}, 
			& \ell = \ell',
			\\
			-e^{-i\Delta_{\boldsymbol{s},j}} (2\cos\Delta_{\boldsymbol{s},j})^{-1}, 
			& \ell = 1- \ell',
		\end{cases}
		\\
		\nonumber
		&\frac{1}{2\pi i} \oint_{\mathscr{C} (\boldsymbol{s}, \ell)} \mathrm{d}z \frac{e^{iE_{\boldsymbol{s}^{[j,j+1]}, \ell'}}}{z-e^{iE_{\boldsymbol{s}^{[j,j+1]}, \ell'}}} \frac{1}{z-e^{iE_{\boldsymbol{s},\ell}}} 
		=
		\left(e^{i(E_{\boldsymbol{s}, \ell}  -E_{\boldsymbol{s}^{[j,j+1]}, \ell'}) } - 1\right)^{-1}
		\\
		&= 
		\begin{cases}
			e^{-i\Delta_{\boldsymbol{s},j,j+1}} (2i\sin\Delta_{\boldsymbol{s},j,j+1})^{-1}, 
			& \ell = \ell',
			\\
			-e^{-i\Delta_{\boldsymbol{s},j,j+1}} (2\cos\Delta_{\boldsymbol{s},j,j+1})^{-1}, 
			& \ell = 1- \ell',
		\end{cases}
	\end{align}
	Therefore, the corrected eigenstates, before taking into account resonances, take the form
	\begin{align}\nonumber
		|\omega_{\boldsymbol{s}, \ell}\rangle &=
		\frac{|\boldsymbol{s},\ell\rangle}{\sqrt{N_{\boldsymbol{s}}}}
		+ \frac{\lambda}{2\sqrt{N_{\boldsymbol{s}}}} \sum_{j=1}^L 
		\left(
		|\boldsymbol{s}^{[j]}, \ell\rangle 
		\frac{\cos(h_js_j)}{\sin\Delta_{\boldsymbol{s}, j}}
		\right.
		\\ \nonumber
		\nonumber
		&
		\left. 
		-|\boldsymbol{s}^{[j]}, 1-\ell\rangle \frac{\sin(h_js_j)}{\cos\Delta_{\boldsymbol{s},j}}
		\right)
		e^{-i\Delta_{\boldsymbol{s},j}}
		\\
		\nonumber
		&+ \frac{\lambda}{2\sqrt{N_{\boldsymbol{s}}}} \sum_{j=1}^L 
		\left(
		|\boldsymbol{s}^{[j,j+1]}, \ell\rangle 
		\frac{\cos(h_js_j+h_{j+1}s_{j+1})}{ \sin\Delta_{\boldsymbol{s}, j,j+1}}
		\right. 
		\\
		&
		\left.
		-|\boldsymbol{s}^{[j,j+1]}, 1-\ell\rangle \frac{\sin(h_js_j+h_{j+1}s_{j+1})}{ \cos\Delta_{\boldsymbol{s},j,j+1}}
		\right)
		e^{-i\Delta_{\boldsymbol{s},j,j+1}} ,
	\end{align}
	where the normalization constant is
	\begin{align}\nonumber
		&N_{\boldsymbol{s}} = 1+ \lambda^2L \left(\bar{V}_1^2+ \bar{V}_2^2\right), 
		\\ \nonumber
		&\bar{V}_1^2 = \frac{1}{4L} \sum_{j=1}^L \left[
		\frac{\cos^2(h_j)}{ \sin^2(\Delta_{\boldsymbol{s},j})}
		+
		\frac{\sin^2(h_j)}{ \cos^2(\Delta_{\boldsymbol{s},j})}
		\right],
		\\
		&\bar{V}_2^2 = \frac{1}{4L} \sum_{j=1}^L \left[
		\frac{\cos^2(h_js_j + h_{j+1}s_{j+1})}{ \sin^2(\Delta_{\boldsymbol{s},j,j+1})}
		+
		\frac{\sin^2(h_js_j + h_{j+1}s_{j+1})}{ \cos^2(\Delta_{\boldsymbol{s},j,j+1})}
		\right].
	\end{align}
	Analogous to Eqs.~\eqref{eq:deg_mat}, we can analyze the resonance effects by diagonalizing within the subspace with non-vanishing matrix elements, including those with a single-spin flip $|\boldsymbol{s},\ell\rangle, |\boldsymbol{s}^{[j]}, \ell'\rangle$ which is the same as Eq.~\eqref{eq:deg_mat},
	and those with two-spin flips  $ |\boldsymbol{s},\ell\rangle, |\boldsymbol{s}^{[j,j+1]}, \ell'\rangle$ can be obtained by changing $\Delta_{\boldsymbol{s},j} \rightarrow \Delta_{\boldsymbol{s},j,j+1}$ and $h_j \rightarrow (h_js_j+ h_{j+1}s_{j+1})$ in Eq.~\eqref{eq:deg_mat}.
	Thus, in parallel to Eqs.~\eqref{eq:xsj} and \eqref{eq:ysj}, we introduce the parameters
	\begin{align}\nonumber
		X'_{\boldsymbol{s},j,j+1} &=
		e^{-i\Delta_{\boldsymbol{s},j,j+1}} 
		\frac{\sin\Delta_{\boldsymbol{s},j,j+1}}{\lambda \cos (h_js_j + h_{j+1}s_{j+1})} 
		\\
		\nonumber
		& \quad
		\cdot \left(
		\sqrt{1+ \left(\frac{\lambda \cos (h_js_j + h_{j+1}s_{j+1})}{\sin\Delta_{\boldsymbol{s},j,j+1}}\right)^2} -1
		\right),
		\\
		\nonumber
		Y'_{\boldsymbol{s},j,j+1} &= 
		-e^{-i\Delta_{\boldsymbol{s},j,j+1}} 
		\frac{\cos\Delta_{\boldsymbol{s},j,j+1}}{\lambda \sin (h_js_j + h_{j+1}s_{j+1})}
		\\
		&\quad 
		\cdot \left(
		\sqrt{1+ \left(\frac{\lambda \sin (h_js_j + h_{j+1}s_{j+1})}{\cos\Delta_{\boldsymbol{s},j,j+1}}\right)^2} -1
		\right).
	\end{align}
	The corrected eigenstates can be written as
	\begin{align}\nonumber
		|\omega_{\boldsymbol{s},\ell}\rangle &= \frac{1}{\sqrt{N_{\boldsymbol{s}}}} \left[ 
		|\boldsymbol{s}, \ell\rangle 
		+ \sum_{j=1}^L \left(
		X_{\boldsymbol{s},j} |\boldsymbol{s}^{[j]}, \ell\rangle 
		+
		Y_{\boldsymbol{s},j} |\boldsymbol{s}^{[j]}, 1-\ell\rangle
		\right)
		\right.
		\\ \nonumber
		&\quad \left. 
		+ \sum_{j=1}^L \left(
		X'_{\boldsymbol{s},j,j+1} |\boldsymbol{s}^{[j,j+1]}, \ell\rangle + Y_{\boldsymbol{s}, j,j+1} |\boldsymbol{s}^{[j,j+1]}, \ell\rangle 
		\right)
		\right] ,
		\\ \nonumber
		&
		N_{\boldsymbol{s}} = 1 + \eta_{\boldsymbol{s}},
		\\
		&
		\eta_{\boldsymbol{s}} =  \sum_{j=1}^L\left( |X_{\boldsymbol{s},j}|^2 + |Y_{\boldsymbol{s},j}|^2 
		+
		|X'_{\boldsymbol{s},j,j+1}|^2 + |Y'_{\boldsymbol{s},j,j+1}|^2\right),
	\end{align}
	which is the result given in Eq.~\eqref{eq:ipr_n2_sec3}.

\end{document}